\input harvmac.tex
\input epsf.tex
\parindent=0pt
\parskip=5pt

\def\ZN{{\bf Z}_N}

\def\IR{{\hbox{{\rm I}\kern-.2em\hbox{\rm R}}}}
\def\IB{{\hbox{{\rm I}\kern-.2em\hbox{\rm B}}}}
\def\IN{{\hbox{{\rm I}\kern-.2em\hbox{\rm N}}}}
\def\IC{{\hbox{{\rm I}\kern-.6em\hbox{\bf C}}}}
\def\IZ{{\hbox{{\rm Z}\kern-.4em\hbox{\rm Z}}}}
\noblackbox

\Title{\vbox{\baselineskip12pt
\hbox{NSF-ITP-96-16}
\hbox{hep-th/9604129}}}
{K3 Orientifolds}

\centerline{ \bf Eric G. Gimon$^a$ and Clifford V. Johnson$^b$}
\bigskip\bigskip\centerline{{\it 
$^a$Department of Physics / $^b$Institute for Theoretical Physics,}}
\centerline{{\it University of California,}}
\centerline{{\it Santa Barbara, CA~93106, USA }}
\footnote{}{\tt $^a$egimon@physics.ucsb.edu, $^b$cvj@itp.ucsb.edu}
\vskip2.5cm
\centerline{\bf Abstract}
\vskip0.7cm
\vbox{\narrower\baselineskip=12pt\noindent
We study string theory propagating on ${\bf R}^6{\times}K3$ by
constructing orientifolds of Type~IIB string theory compactified on
the $T^4/{\bf Z}_N$ orbifold limits of the $K3$ surface. This
generalises the ${\bf Z}_2$ case studied previously. The orientifold
models studied may be divided into two broad categories, sometimes
related by T--duality.  Models in category $A$ require {either}
both D5-- and D9--branes, { or} only D9--branes, for consistency.
Models in category $B$ require { either} only D5--branes, { or}
no D$-$branes at all.  This latter case is an example of a consistent
purely closed unoriented string theory.  The spectra of the resulting
six dimensional ${\cal N}{=}1$ supergravity theories are presented.
Precise statements are made about the relation of the ${\bf Z}_N$ ALE
spaces and D5--branes to instantons in the dual heterotic string theory.}

\vskip0.5cm

\Date{April, 1996 (corrected 9th July '96)}

\baselineskip13pt

\def\Z{{\bf Z}}
\lref\dbranes{J.~Dai, R.~G.~Leigh and J.~Polchinski, Mod.~Phys.~Lett.
{\bf A4} (1989) 2073\semi
P.~Ho\u{r}ava, Phys. Lett. {\bf B231} (1989) 251\semi
R.~G.~Leigh, Mod.~Phys.~Lett. {\bf A4} (1989) 2767\semi
J.~Polchinski, Phys.~Rev.~D50 (1994) 6041, hep-th/9407031.}
\lref\ericjoe{E. G. Gimon and J. Polchinski, {\sl `Consistency 
Conditions for Orientifolds and D--Manifolds'}, hep-th/9601038.}
\lref\atish{A. Dabholkar and  J. Park, {\sl `An Orientifold of 
Type--IIB Theory on $K3$'}, hep-th/9602030.}
\lref\gojoe{J. Polchinski, {\sl `Dirichlet Branes and Ramond--Ramond
Charges in String Theory.'}, Phys. Rev. Lett. {\bf 75} (1995) 4724,
hep-th/9510017.}
\lref\polcai{J. Polchinski and Y. Cai, Nucl. Phys. {\bf B296} (1988) 91.}
\lref\curt{C. G. Callan, C. Lovelace, C. R. Nappi and S.A. Yost,
Nucl. Phys. {\bf B308} (1988) 221.}
\lref\dnotes{J. Polchinski, S. Chaudhuri and C. V. Johnson, {\sl 
`Notes on D--Branes'}, hep-th/9602052.}
\lref\opentwist{A. Sagnotti, in {\it
Non-Perturbative Quantum Field Theory,}
eds. G. Mack et. al. (Pergamon Press, 1988) 521;\hfil\break
M. Bianchi and A. Sagnotti, Phys. Lett. {\bf 247B}
(1990) 517; Nucl. Phys. {\bf B361} (1991) 519;\hfil\break
P. Horava, Nucl. Phys. {\bf B327} (1989) 461;
Phys. Lett. {\bf B289} (1992) 293; Nucl. Phys. {\bf B418}
(1994) 571; {\it Open Strings from Three Dimensions: Chern-Simons-Witten
Theory an Orbifolds,} Prague preprint PRA-HEP-90/3, to appear in J. Geom.
Phys. }
\lref\openstuff{J. A. Harvey and J. A. Minahan, Phys. Lett. {\bf 188B}, 
44 (1987)\semi
P. Horava, Nucl. Phys. {\bf B327} (1989) 461\semi
N. Ishibashi and T. Onogi, Nucl. Phys. {\bf B318} (1989) 239\semi
P. Horava, Phys. Lett. {\bf B231}, 251 (1989)\semi
Z. Bern, and D. C. Dunbar, Phys. Lett. {\bf B242} (1990) 175\semi
Phys. Rev. Lett. {\bf 64} (1990) 827\semi
Nucl. Phys. {\bf B319} (1989) 104; Phys. Lett. {\bf 203B} (1988) 109\semi

G. Pradisi and A. Sagnotti, Phys. Lett. {\bf B216} (1989) 59\semi
M. Bianchi and A. Sagnotti, Phys. Lett. {\bf B247} (1990) 517\semi
Nucl. Phys. {\bf B361} (1991) 519\semi
A. Sagnotti, Phys. Lett. {\bf
B294} (1992) 196\semi
A. Sagnotti, {\sl `Some Properties of Open--String
Theories'}, preprint ROM2F-95/18, hep-th/9509080.
}
\lref\stringstring{M. J. Duff and R. R. Khuri, 
Nucl. Phys. {\bf B411} (1994) 473\semi
M. J. Duff and R. Minasian, Nucl. Phys. {\bf B436} (1995) 507.}
\lref\ed{E.~Witten, {\sl `String Theory Dynamics in Various
 Dimensions'}, 
Nucl. Phys. {\bf B443} (1995) 85, hep-th/9503124.}
\lref\dab{A. Dabholkar,
Phys. Lett. {\bf B357} (1995) 307.}
\lref\hull{C. M. Hull,
Phys. Lett. {\bf B357} (1995)
545.}
\lref\edjoe{J. Polchinski and E. Witten {\it Evidence for Heterotic 
--- Type I Duality,}
preprint IASSNS-HEP-95-81, hep-th/9510169.}
\lref\sagnotti{A. Sagnotti, {\sl `A Note on the
 Green--Schwarz Mechanism in Open--String Theories'},
 Phys. Lett. {\bf B294} (1992) 196.} 
\lref\solitons{A. Strominger {\sl `Heterotic Solitons'},
 Nucl. Phys. {\bf B343} (1990) 167\semi
C. G. Callan, J. A. Harvey and A. Strominger, {\sl `Solitons 
in String
Theory',}, Trieste Notes, World Scientific (1991).}

\lref\small{E. Witten, {\sl `Small Instantons in String
Theory',} preprint IASSNS-HEP-95-87, hep-th/9511030.}
\lref\joetc{M. Berkooz, R. G. Leigh, J. Polchinski, J. H. Schwarz and N. 
Seiberg, {\sl `Anomalies and Dualities in Six--Dimensional Type I
Superstrings'}, to appear.}

\lref\eguchi{T. Eguchi and A. J. Hanson, {\sl `Asymptotically 
Flat Self--Dual Solutions To Euclidean Gravity'}, Phys. Lett. {\bf B74} 
(1978), 249.}
\lref\gibbons{G. W. Gibbons and S. W. Hawking, {\sl `Gravitational 
Multi--Instantons'}, Phys. Lett. {\bf B78}, (1978), 430.}

\lref\town{P. K.~Townsend, {\sl `P--brane Democracy'},  
hep-th/9507048.}
\lref\conif{A. Strominger,  Nucl. Phys. {\bf B451} (1995) 96, 
hep-th/9504090 }
\lref\jkkm{C. V. Johnson, N. Kaloper, R. R. Khuri and R. C. Myers, {\sl `Is
String Theory a Theory of Strings?'}, Phys. Lett. {\bf B368} (1996)
71, hep-th/9509070.}

\lref\page{D. Page, Phys. Lett. {\bf B80} (1978) 55.}
\lref\green{M. B. Green and J. H. Schwarz, Phys. Lett. {\bf B149} (1984) 117.}
\lref\walton{M. A. Walton, Phys. Rev. {\bf D37} (1988) 377.}
\lref\andycumrun{A. Strominger and C. Vafa, {\sl `Microscopic 
Origin of the Bekenstein--Hawking Entropy'}, hep-th/9601029. }
\lref\edvafa{C. Vafa and E. Witten, {\sl `Dual String Pairs With $N{=}1$ 
And $N{=}2$ Supersymmetry In
Four Dimensions'}, Nucl. Phys. {\bf B447}
(1995) 261, hep-th/9507050.}
\lref\thetabook{H. E. Rauch and A. Lebowitz, {\sl `Elliptic 
Functions, Theta Functions, and Riemann Surfaces'}, Williams and  Wilkins
 (1973).}
\lref\seiberg{N. Seiberg,  Nucl. Phys. {\bf B303}, (1988) 286.}
\lref\mikegreg{M. R. Douglas and G. Moore, {\sl `D--branes, Quivers, 
and ALE Instantons'}, hep-th/9603167.}
\lref\hethet{M. J. Duff, R. Minasian and E. Witten,
{\sl `Evidence for Heterotic/Heterotic Duality'}, hep-th/9601036.}
\lref\moretocome{E. Gimon and C. V. Johnson, to appear.}
\def\pkm{\!\textstyle{{\pi k\over N}}\,}
\def\epkm{\textstyle{\left({2\pi k\over N}\right)}}
\def\opkm{\!\textstyle{{(2k-1)\pi\over N}}\,}
\newsec{Introduction}
A significant advance in the ability of string theorists to calculate
in both perturbative and non--perturbative string theory was made when
the link between D--branes\dbranes\ and Ramond--Ramond (R-R) states in
types I and II string theory was established\gojoe. The precise
statement is that D--branes, extended objects with
$(p+1)$--dimensional world volume\foot{We shall hereafter refer
specifically to the $p$--dimensional objects as `D$p$--branes'. This
will allow us to reserve terms like `D--brane' for the generic case.}\
are the natural carriers of the fundamental units of $(p+2)$--form R-R
field strength charge.  Furthermore, D--branes are
Bogomol'nyi--Prasad--Sommerfeld (BPS) saturated states, the very
states about which the most far--reaching statements in
non--perturbative string theory have been made.  The traditional
string theory toolbox is well equipped to study many aspects of
D--branes\foot{For a review, see ref.\dnotes.}.  Consequently, many of
the conjectures and partial computations which were made concerning
non--perturbative string theory, (including suggestions that other
extended objects were unavoidable outside perturbation
theory\refs{\town,\jkkm,\conif}) were able to be tested and improved
using D--brane technology.

Of the recent developments in string theory, `duality' has been been
the physical phenomenon receiving most of the attention, due to the
wide range of results and ideas it organises.  One of the most
striking manifestations of duality has been the idea that a string
theory in its strong coupling limit yields another --- weakly coupled
--- string theory\stringstring\ed.  Specifically, each of the known
superstring theories have been demonstrated (sometimes after a
compactification) to be related to another by some type of
strong--weak coupling duality. A result of this has been that the
previous emphasis (motivated by phenomenology) on heterotic string
theory has become less pronounced: It is now believed that the known
superstring theories are different perturbative realisations of some
more profound underlying theory. In order to understand the nature of
this unknown theory it seems prudent to gather as much data about it
as possible, by studying more examples of stringy vacua which will
give us insight into duality and related issues.

One such string--string duality pair is given by the $SO(32)$ `type~I'
and `heterotic' string theories. As ten dimensional theories, they
were conjectured to be strong--weak coupling duals in ref.\ed. The
ideas presented there, together with the later computations in
refs.\refs{\dab,\hull}, were lent considerable support by the
convincing computations (involving D--branes) carried out in
ref.\edjoe.  A natural step in studying this duality family further is
to study non--trivial compactifications of both theories which are
dual to one another. An immediate consequence of such a study was the
realisation that there can be non--perturbative contributions to the
gauge group of $SO(32)$ heterotic string theory (a phenomenon
previously noted only in type~II string theory\ed\conif), which have a
perturbative origin in the type~I theory\small. The observed
non--perturbative phenomena were identified with the zero--size limit
of heterotic instantons, which are ordinary solitonic
fivebranes\solitons\ in heterotic string theory ({\sl i.e.,} they give
order $e^{-1/\lambda^2_h}$ effects) and Dirichlet fivebranes (order
$e^{-1/\lambda_I}$) in the type~I theory.  In the type~I theory, these
fivebranes are groups of D5--branes which are constrained to move as a
single unit.  The simplest such object is a pair of D5--branes,
carrying $SU(2)$ Chan--Paton factors, as conjectured in ref.\small.
An immediate consequence of this conjecture was that a family of $2M$
D5--branes has gauge group $U\!Sp(2M)$. This corresponds to a
non--perturbative contribution to the gauge group of the heterotic
string, where it arises as $M$ small instantons coincide.

Further investigations into such dual phenomena, and involving other
compactifications of the dual theories, require more understanding of
perturbative type I compactifications. The technology of such
compactifications is perhaps the least developed of all of the
superstring theories, probably due to the fact that open and
unoriented string theories are not fully constrained by modular invariance
of the worldsheet theory, in contrast to the closed oriented
cases. The consistency of such compactifications traditionally
involves computations which may be interpreted as the cancelation of
tadpoles\refs{\polcai,\curt}\ and the removal of spacetime anomalies.
This is still true at the present time.

The extension of the consistency techniques to include more general
D--brane configurations was presented quite recently in
ref.\ericjoe. A perturbative type~I derivation of the aforementioned
symplectic projection in the D5--brane sector was presented there,
confirming the conjecture and subsequent results of ref.\small.  In
addition, ref.\ericjoe\ presented an example of a non--trivial
compactification: The type~I string on $K3$ in its ${\bf Z}_2$
orbifold limit.  This example is interesting for a number of reasons.
One of these is the appearance of a new mechanism for the appearance
of enhanced gauge symmetry groups: The basic dynamical fivebrane in
the models of \ericjoe\ is composed of four D5--branes, contributing
$SU(2)$ to the gauge group when in isolation and away from the fixed
points of the orbifold. The number four arises because the ${\bf Z}_2$
action on flat spacetime is realised on at least two D5--branes, while
the orientation projection (by which we obtain type~I strings from
type~II strings) requires a further doubling. We can therefore think
of the four--D5--brane object in this case as two (mirror image)
copies of the two--D5--brane unit discussed in ref.\ed.

In the models of ref.\ericjoe, the enhancement of the gauge symmetry
to $U\!Sp(2M)$ when $M$ of these dynamical fivebranes coincide occurs
in the same way as in the ten dimensional theory. However there is
extra structure in the $K3({\bf Z}_2)$ case, due to the presence of
the sixteen orbifold fixed points. When $2M$ D5--branes coincide at
one of these fixed points, there is an enhanced gauge symmetry $U(M)$,
an interesting new feature.  Another interesting observation is the
fact that some of the models arising in this type~I $K3({\bf Z}_2)$
compactification have interesting heterotic duals\joetc.  The gauge
group has a contribution from D9--branes and D5--branes.  Generalising
the results of \small, on the heterotic side the origin of this gauge
group is therefore viewed as having both perturbative and
non--perturbative components, in the dual heterotic string theory.
The details of this have been worked out in ref.\joetc. Also discussed
there is the interesting fact that the spectra which arise have been
shown to have a new class of $U(1)$ anomalies for which the anomaly
polynomial does not factorise. These anomalies can be shown to be
canceled by a generalisation of the standard Green--Schwarz
mechanism\green\sagnotti\joetc\atish.
 
In this paper we present further studies of the type~I $K3$
compactifications, considering other orbifold limits\foot{See also
ref.\openstuff\ for more work in this context.}.  Specifically, our
focus here will be to apply the consistency constraints presented in
ref.\ericjoe, in the case where the orientifold group (with which we
construct type~I theories from type~IIB) contains $\ZN$ as its
spacetime symmetry group.  The models arising are a natural extension
of the models of \ericjoe. One might expect that they will shed
further light upon the nature of the types of symmetry enhancement
possible at the orbifold points of the $K3(\ZN)$ models, and this is
indeed the case. For $K3(\ZN)$ the basic dynamical unit is now $2N$
D5--branes; $N$ for a natural action of $\ZN$ in spacetime on the
D5--branes, together with the pairing in the orientifold. The
coincidence of these dynamical fivebranes results in symplectic gauge
groups as before, with enhancement to unitary groups at the orbifold
fixed points.  Particularly interesting is the fact that for ${\bf
Z}_4$ and ${\bf Z}_6$, the types of fixed points are mixed, which
makes for extra non--trivial structure which helps us to deduce
certain facts about the nature of open string theory near the $\ZN$
ALE spaces in their blow--down limit.

After a brief review in section~2 of the properties of $K3$'s $\ZN$
orbifold limits which we will use, we study in section~3 how the
action of $\ZN$ orientifold groups are realised at the level of
Chan--Paton factors. In section~4, we present detailed computation of
the twisted sector tadpoles which arise in the neighbourhood of the
fixed points of an orbifold. This supplies us with useful information
about the nature of type I string theory on the zero size limit of
$\ZN$ ALE instantons, out of which a $K3$ manifold may be constructed
in the standard way.  In section~5 we put all this information
together and compute the complete spectra of type~IIB string theory on
$K3$ orientifolds. Throughout, we discuss and develop results as they
arise, concluding in section~6 with a brief summary and
discussion. The data we gained about these models will have a lot to
teach us about the dual heterotic string theory, and we briefly
interpret our results in this light.

\newsec{ALE Spaces, Orbifolds  and $K3$}
We will study string propagation on ${\bf R}^6{\times}K3$, denoting by
$X^\mu$, $\mu=0,1,\ldots,5$, the non--compact coordinates and $X^m$,
$m=6,\ldots,9$ the compact coordinates. For most of the presentation
we will consider $K3$ in its orbifold limits, $T^4/\ZN$. To construct
the orbifold $T^4/\ZN$, we begin with the space ${\bf R}^4\equiv{\bf
C}^2$, with complex coordinates $z^1=X^6+iX^7$ and $z^2=X^8+iX^9$, upon
which we make the identifications $z^i\sim z^i+1\sim z^i+i$, for
$N{=}2$~or~4, and $z^i\sim z^i+1\sim z^i+\exp({\pi i/3})$ for
$N{=}3$~or~6.  These lattices define for us the torus $T^4$, upon which
the discrete rotations ${\bf Z}_N$, acts naturally as
$(z^1,z^2)\to(\beta z^1,\beta^{{-}1}z^2)$, for $\beta=\exp(2\pi i/N)$.

We may therefore define a new space by identifying points under the
action of $\ZN$. This is the orbifold $T^4/\ZN$, which is a smooth
surface except at points at which the curvature of the orbifold is
located. These `fixed points' are points which are invariant under
$\ZN$ or some non--trivial subgroup of it.  For $N\in\{2,3,4,6\}$,
this procedure produces a family of compact spaces which are the
orbifold limits of the $K3$ surface.

The smooth $K3$ manifold is obtained from these limits by `blowing up'
the orbifold points. This procedure is simply the process by which
each of the points is removed and replaced by a smooth space. For the
orbifold $T^4/\ZN$ the neighbourhood of a fixed point is ${\bf
R}^4/{\bf Z}_M$, where $N\geq M\in\{2,3,4,6\}$. This is the asymptotic
region of the A--series ALE gravitational 
instanton\refs{\eguchi,\gibbons}\ 
(denoted here
${\cal E}_M$), with which we replace the excised point.

Denote the generator of $\ZN$ by $\alpha^{\phantom{m}}_N$, the group
elements being the powers $\alpha_N^m$, for $m\in\{0,1,\ldots,N-1\}$.
First, we observe that the number, $F_M$, of points fixed under the
${\bf Z}_M$ subgroup of $\ZN$, (generated by $\alpha^{N/M}_N$) is
simply $F_M=16\sin^4{\pi\over M}$, where $M$ is a multiple
of $N$.

Let us review the fixed point structure of each space. For more
details, see refs.\refs{\page,\walton}. For $T^4/{\bf Z}_2$ we have 16
points fixed under the action of $\alpha_2$, each of which are
replaced by the space ${\cal E}_2$, in order to resolve to smooth
$K3$.  Meanwhile for $T^4/{\bf Z}_3$ there are 9 fixed points of
$\alpha_3$, which are each replaced by ${\cal E}_3$ in the blow--up.

The case $T^4/{\bf Z}_4$ has 16 fixed points. Four of them are fixed
under the action of $\alpha^{\phantom{2}}_4$,
while the other 12 are only fixed under $\alpha^2_4$. 
Under $\alpha^{\phantom{2}}_4$,
these 12 ${\bf Z}_2$ points transform as 6 doublets.  Consequently, the
blow--up is carried out by first constructing the ${\bf
Z}_4$--invariant region by identifying these pairs of fixed points, and
then replacing each of the original 4 ${\bf Z}_4$ fixed points by an
${\cal E}_4$ and the 6 pairs  by an ${\cal E}_2$.

For $T^4/{\bf Z}_6$ the situation is similar. There are 24 fixed
points altogether.  There is only one point fixed under
$\alpha^{\phantom{3}}_6$. It is replaced by ${\cal E}_6$ in the
blow--up. There are 8 points fixed under the ${\bf Z}_3$ subgroup,
generated by $\alpha^2_6$, which transform as doublets under the
action of $\alpha^{\phantom{3}}_6$. They are therefore replaced by 4
copies of ${\cal E}_3$. There are 15 points fixed under $\alpha^3_6$,
which transform as triplets under the action of
$\alpha^{\phantom{3}}_6$. Consequently, they are replaced by 5 copies
of ${\cal E}_2$ in performing the blow--up surgery.

Knowledge of the properties of the ALE spaces ${\cal E}_m$ which are
used in the blow--up procedure can be used to show that many of the
properties of $K3$ can be deduced from this construction\page.

\newsec{Orientifolds and Chan--Paton Factors}
We wish to consider type I string theory propagating on the spaces we
described above. Such theories are constructed here by orientifolding
type~IIB string theory and introducing open string boundary
conditions. An orientifold generalises the concept of an orbifold to
include not only spacetime symmetries, but also world sheet parity
symmetry in the group of discrete symmetries which are gauged.  The
introduction of open string sectors is somewhat analogous to
introducing twisted sectors in an orbifold theory\opentwist.

Our spacetime symmetry group is $\ZN$. As before, we denote the
generator of this group by $\alpha^{\phantom{k}}_N$, the group
elements being the powers $\alpha_N^k$, for
$k\in\{0,1,2,\ldots,N-1\}$.  The orientifold group can therefore
contain the elements $\alpha_N^k$ and also $\Omega\cdot\alpha^k_N$
(which we shall sometimes denote $\Omega_k$), where $\Omega$ is world
sheet parity.  Gauging the action of $\alpha_N^k$ will introduce the
familiar closed string twisted sectors for an orbifold, while gauging
$\Omega_k$ will result in unoriented surfaces in the world--sheet
expansion.

We have a choice as to the elements which we wish to consider in our
orientifold group, our only constraint being closure, of course. Let
us denote the two choices of $\ZN$ orientifold group as $\ZN^A$ and
$\ZN^B$. The choice analogous to ref.\ericjoe\ is to have
\eqn\orienta{\ZN^A=\{1,\Omega, \alpha^k_N,\Omega_j\},
\quad k,j=1,2,\ldots N-1.} A second choice (only for $N$ even) is
\eqn\orientb{\ZN^B=\{1, \alpha^{2k-2}_N,\Omega_{2j-1}\},
\quad k,j=1,2,\ldots {N\over2}.} Both of these orientifold groups
are consistent, as they both close under group multiplication. We
shall see the consequences of each choice of orientifold group as we
proceed.
\def\ip{i^\prime}
\def\jp{j^\prime}

Consistency will require the addition of open string sectors\opentwist. Open
string endpoints will lie on various submanifolds of spacetime, the
D--branes.  Each such submanifold is labeled by a Chan-Paton index
corresponding to the state of an open string endpoint.  An open string
state will be denoted $|\psi,ij\!>$, where $\psi$ is the state of the
world--sheet fields and $i$ and $j$ are the states of the string
end--points. For a consistent construction, we must determine how the
action of the orientifold group is manifested at the level of
Chan--Paton factors.  In general, for every D--brane which exists in
the theory, the spacetime transformed D--brane must also
appear\ericjoe. The action of an orbifold group element $g$ on this
complete set will be represented by the unitary matrices $\gamma_g$
which act on the open string endpoints.

Constraints on the $\gamma$'s arise when we consider
the action of various orientifold symmetries: They must form a
faithful representation of the group, up to phases.  We have for example 
\eqn\actionthree{\alpha_N^k:\quad |\psi,ij\!>\quad\to\quad
 (\gamma^{\phantom{-}}_k)_{i\ip}
|\alpha_N^k\cdot\psi,\ip\jp\!>(\gamma^{-1}_k)_{\jp j}} 
while for $\Omega\cdot\alpha_N^k\equiv\Omega_k$,
\eqn\actionfour{\Omega_k:\quad |\psi,ij\!>\quad\to\quad 
(\gamma^{\phantom{-1}}_{\Omega_k})_{i\ip} |\Omega_k
\cdot\psi,\jp\ip\!>(\gamma^{-1}_{\Omega_k})_{\jp j}.}   
Notice that when that action
includes $\Omega$,  the ends of the string are
transposed.
Composing various actions of the group elements, we see that as 
$(\alpha_N^k)^N=1$, then 
\eqn\actionfive{(\alpha_N^k)^N:\quad |\psi,ij\!>\quad\to\quad
 (\gamma^{N}_k)_{i\ip}
|\psi,\ip\jp\!>(\gamma^{-N}_k)_{\jp j}} and so 
\eqn\actionsix{\gamma_{k}^N=\pm1.}
Similarly, as $\Omega^2=1$
\eqn\actionseven{\Omega^2:\quad |\psi,ij\!>\quad\to\quad 
(\gamma^{\phantom{-1}}_{\Omega}(\gamma^{T}_{\Omega})^{-1}
)_{i\ip}
|\psi,\ip\jp\!>(\gamma^{T}_{\Omega}\gamma^{-1}_{\Omega})_{\jp
j},}
resulting in
\eqn\resultone{\gamma^{\phantom{-1}}_{\Omega}
=\pm\gamma^{T}_{\Omega}.}  
Other examples of such
conditions will be put to explicit use later when solving the tadpole
equations. Another important example of a physical constraint is the
`superselection rule' which implies that there should be no non--zero
elements of the Chan--Paton matrices, $\lambda$, which connect
D--branes which are at different points in spacetime.

\newsec{Tadpoles for ALE ${\bf Z}_M$ Singularities.}
\subsec{Consistent Field Equations}
In open and/or unoriented string theory, certain divergences arise at
the one--loop level, which may be interpreted\polcai\ as
inconsistencies in the field equations for the R-R potentials in the
theory. They manifest themselves as tadpoles.  A minimum requirement
in the construction of a consistent theory is that we ensure that
these tadpoles are all canceled. These tadpoles are topologically of
two types, disc tadpoles and ${\bf RP}^2$ tadpoles. They are perhaps
best visualised as the process of emitting an R-R closed string state
from a D$p$--brane (for the disc), a source of $(p+1)$--form R-R
potential, or from an orientifold plane (for ${\bf RP}^2$), which also
carries R-R charge.  The prototype example of the cancelation of these
tadpoles is the ten  dimensional $SO(32)$ open (type~I) string
theory. In that case the disc and ${\bf RP}^2$ produce a divergence
proportional to $(n_9{-}32)^2$ for $SO(n_9)$ Chan--Paton factors ({\sl
i.e.}, there are $n_9$ D9--branes\gojoe) and $(n_9{+}32)^2$ for
$USp(n_9)$. Cancelation of the divergences therefore requires gauge
group $SO(32)$ ({\sl i.e.}, 32 D9--branes). In that case, the
orientifold group was very simple, the only element being the purely
internal $\Omega$, and thus there are no spacetime symmetries to
consider. The tadpole cancelation there was for global consistency of
the 10--form potential's field equation.
 
In our case, we have all of the spacetime symmetries $\ZN$ to
include. So in general, we will have tadpole contributions from not
only the familiar R-R potentials, but also twisted R-R potentials. For
orientifold group $\ZN^A$, which contains $\Omega$, we will have
D9--branes present, as in ten dimensional type~I string theory and in
the ${\bf Z}_2$ example of ref.\ericjoe. In the case of $\ZN^B$, there
is no element $\Omega$, and we will have no requirement to include
D9--branes, as in the example of \atish.  In the case of type $A$
orientifolds we will always have D5--branes whenever $N$ is even.
This is simply because only for $N$ even does there exist a ${\bf
Z}_2$ subgroup of ${\bf Z}_N$. There are many ways to see that this
results in D5--branes.  It will be particularly obvious when we
consider the tadpole diagrams later.

One  way to think about it is to realise that the ${\bf Z}_2$
subgroup acts as reflections in the $X^m$, $(m=6,\ldots,9)$
directions, generated by $\alpha^{N/2}_N$ (called $R$ in
ref.\ericjoe). Therefore there is also the element
$\Omega_{N/2}\equiv\Omega R$ in the orientifold group.  Under
$T$--duality in the directions in which $R$ acts, $\Omega R$ becomes
$\Omega$. Whatever the details of the dual model, it must contain
D9--branes, because of the presence of $\Omega$.  T--duality exchanges
Dirichlet  (D) and Neumann (N) boundary conditions. Therefore, the original
model with orientifold group containing the element $\Omega R$
contains D5--branes, with $X^m$ as their Dirichlet coordinates, their
world--volumes filling the non--compact ${\bf R}^6$.  

\def\ZM{{\bf Z}_M}
Henceforth, we shall use ${\bf Z}_M$ for the subgroups of the $\ZN$ we
use for the orbifold, where $M\leq N$ is a factor of $N$. This will be
used to denote the generic subgroup under which the orbifold
singularities (with local ALE geometry ${\bf R}^4/{\bf Z}_M$) are
fixed.  ($N$ will be reserved for the parent group).  D--branes and
fixed points are sources for R-R field strengths.  In particular, a
D5--brane sitting at a $\ZM$ fixed point can be a source for various
$\alpha^m_M$--twisted 6--form R-R potentials, $A^m_6$, with field
strength $H_7^m$, where $m=1,2,\ldots M-1$.  The open string one--loop
diagram corresponding to the twisted cylinder shows two D5--branes
exchanging such a form in the closed string channel:
 
\vskip1.0cm
\hskip5cm\epsfxsize=1.2in\epsfbox{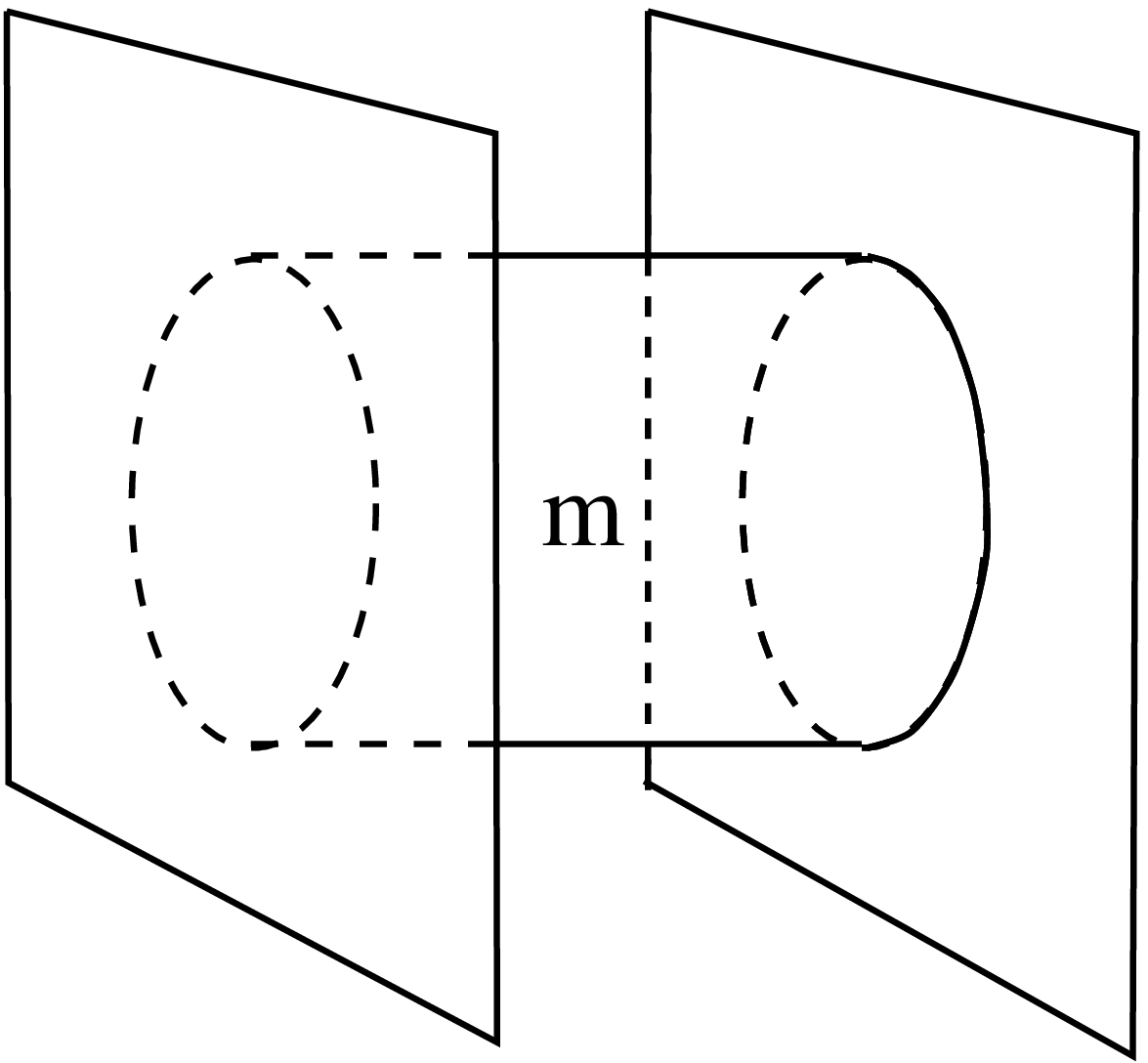}

Let us denote the charge of the D5--brane by $\mu_5^m$.
A crosscap diagram arises in the
neighbourhood of a fixed point, which is also an orientifold point.
The orientifold/fixed point identifies strings under $\Omega$ together
with a phase from the group element $\alpha^{k}_M$. This is the
orientifold group element $\Omega_k$: fixed points can have charge
under the $A^m_6$, when $m=2k$. 
The open string amplitude depicting the interaction between a
D5--brane the fixed point is a M\"obius strip:
 
\hskip5cm\epsfxsize=1.2in\epsfbox{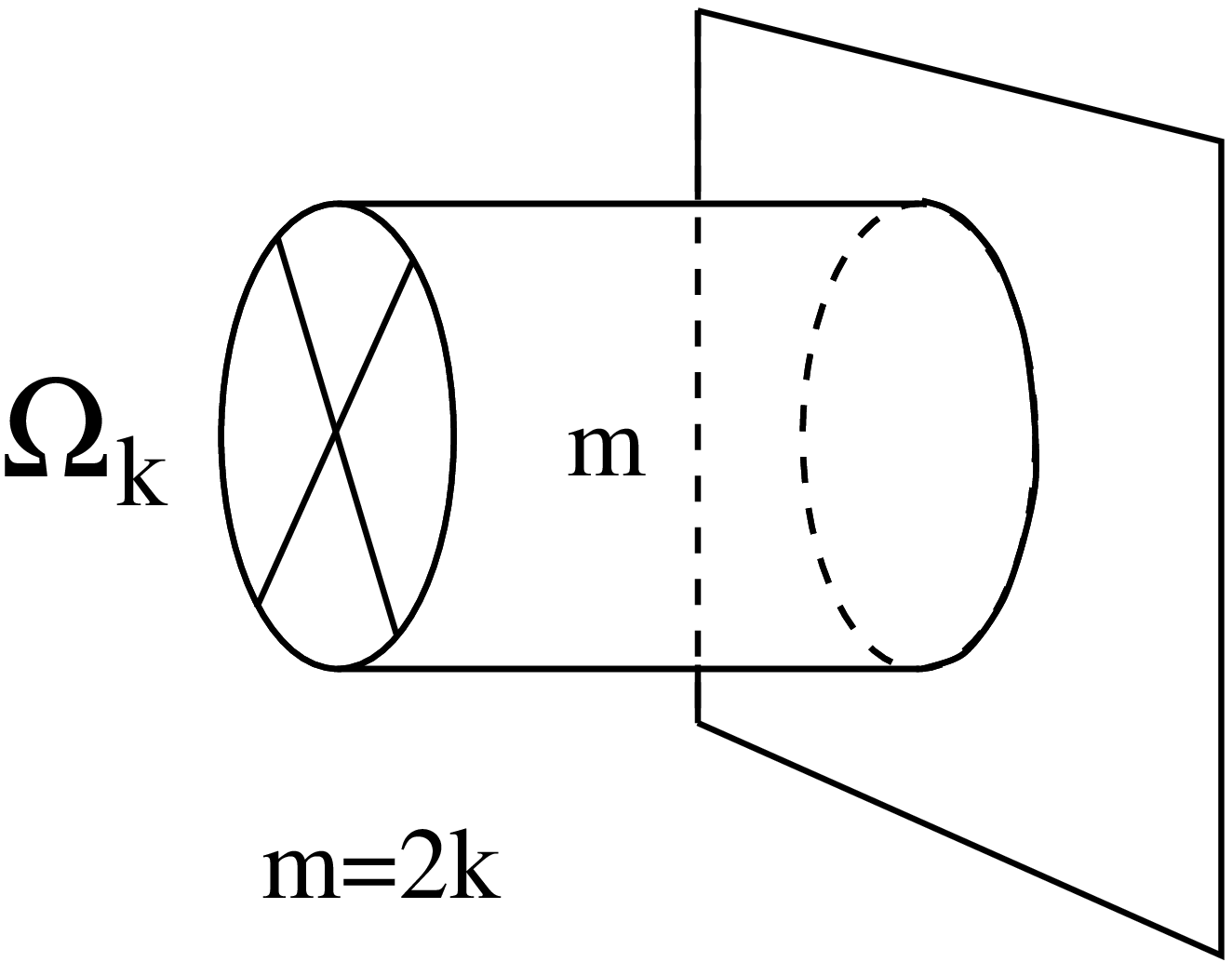}

From this amplitude, one may formally extract the value of the force
between the D5--brane and the fixed point, as in ref.\dnotes. By
comparing it to the force between two D5--branes, obtained from the
cylinder amplitude we may deduce the charge ${\tilde\mu}^m_5$ of the
fixed point.  Note, however that for non--trivial twists by $m$, the
string zero modes vanish, and therefore the fivebrane is forced to sit
at the fixed point.  This simply means that the twisted 6--form R-R
potentials $A_6^m$ cannot propagate in spacetime. Therefore, their
equations of motion must be satisfied locally.  It is only when we
consider the untwisted R-R six--form field strength $A_6$ ({\sl i.e.,}
$m{=}0\to k{=}M/2$ mod $M$) that we see how to separate the D5--branes
from the fixed points. As we shall see, there will be freedom in the
final equations to group together a collection of D5--branes, forming
them into a collective object with $H_7^m$ charge ${\mu}^m_5=0$, but
non--zero charge $\mu_5$ under the untwisted field strength,
$H_7$. This fivebrane is now free to move away from fixed point; it is
the generalisation of the dynamical fivebrane of refs.\small\ and
\ericjoe. In the case of ref.\small, the fivebrane (in the ten
dimensional theory) was composed of two D5--branes, while on $K3({\bf
Z}_2)$ it was composed of four of them\ericjoe.  In the present
$K3(\ZN)$ context, we shall see that the dynamical fivebrane is
composed of $2N$ D5--branes (for orientifold group of type $A$). As
the fivebrane, charged under $H_7$, can move about in the internal
space, the conservation equations for $A_6$ charge need not be
canceled locally. Flux lines can extend throughout the compact
internal space, the only requirement being global charge conservation.
In the case when the orientifold group is $\ZN^A$, we also have the
familiar ten--form potential's field equations to satisfy globally.

The way which we shall proceed is as follows. We shall compute all of
the tadpoles arising from the various orientifold group elements which
might appear, focusing our attention initially on the twisted 6--form
potentials' tadpoles. We can therefore carry out a local analysis, in
the neighbourhood of a fixed point.  Therefore, we can simply consider
our `internal' space to be ${\bf R}^4/{\bf Z}_N$, (the zero--size
`blow--down' limit of an ALE instanton). Our `internal' space is
therefore non-compact, for the purposes of this local
computation.  We can therefore ignore most of the clutter of the
propagation of zero modes and the presence of winding modes in the
`internal' directions, as these will not be relevant in the
limit.  The tadpoles which arise from the untwisted 6--form  and
10--form potentials will be identical to those already computed in
ref.\ericjoe. (We will nevertheless compute and present them here, for
completeness.)

The information we will gather about local
tadpole cancelation at an ALE point can then be used to construct the
complete $K3$ orbifold model, by using the knowledge we have about how
$K3$ is constructed from such points (reviewed in section~2). 

\subsec{Preliminaries}
The most efficient way of computing the divergent contribution of the
tadpoles is to compute the one--loop diagrams (the Klein bottle (KB),
M\"obius strip (MS) and cylinder (C)) and then to take a limit which
extracts the divergent pieces. The fact that these diagrams yield the
disc and ${\bf RP}^2$ tadpoles in terms of the sums of three different
products means that the requirement of factorisation of the final
expression is a strong consistency check on the whole computation.

The three consistent types of diagram
which can be drawn, labeled by the possible elements of the
orientifold group under consideration are depicted below: 

\vskip0.5cm
\hskip0.5cm\epsfxsize=4.0in\epsfbox{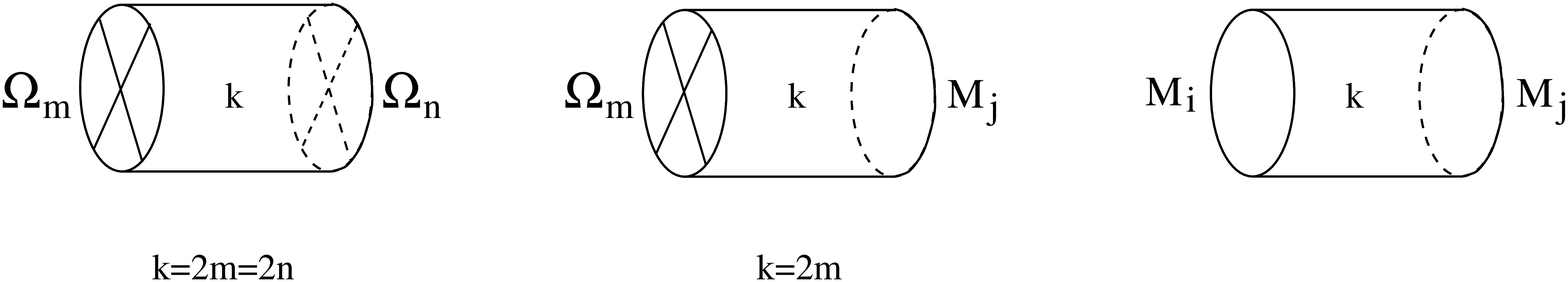}
\vskip0.1cm

In the figure, the crosscaps show the action of $\Omega_m$ as one goes
half way around the open string channel ({\sl around} the
cylinder). Going around all the way picks up another action of
$\Omega_m$, yielding the $\ZN$ element $\Omega^2_m=\alpha^{2m}_N$
which is the twist which propagates in the closed string channel ({\sl
along} the cylinder).  For consistency, if there is a crosscap with
$\Omega_n$ at the other end, forming a Klein bottle, then
$\Omega^2_n=\alpha^{2n}_N$ should yield the same twist in the closed
string channel, {\sl i.e.}, $2n=2m$.  For $\ZN$ with $N$ odd, there is
only one solution to this: $m=n$ mod $N$.  When $N$ is even however,
we can have also the solution $n=m+N/2$ mod~$N$.  Note also that in
the figure, $ M_i$ denotes the manifold $i$, upon which an open string
can end: a D--brane.

Let us parameterise the surfaces as cylinders with length $2\pi l$ and
circumference $2\pi$ with either boundaries or crosscaps on their ends
with boundary conditions on a generic field $\phi$ (and its
derivatives):
\eqn\paramet{\eqalign{{\rm KB}:\quad&\phi(0,\pi+\sigma^2)
=\Omega_m\cdot\phi(0,\sigma^2)\cr
&\phi(2\pi l,\pi+\sigma^2)=\Omega_n\cdot\phi(2\pi l,\sigma^2)\cr
&\phi(\sigma^1,2\pi+\sigma^2)=\alpha^k_N\cdot\phi(\sigma^1,\sigma^2);
\quad k=2m=2n\cr
{\rm MS}:\quad&\phi(2\pi l,\sigma^2)\in M_j\cr
&\phi(0,\pi+\sigma^2)=\Omega_m\cdot\phi(0,\sigma^2)\cr
&\phi(\sigma^1,2\pi+\sigma^2)=\alpha^k_N\cdot\phi(\sigma^1,\sigma^2);
\quad k=2m\cr
{\rm C}:\quad&\phi(0,\sigma^2)\in M_i\cr
&\phi(2\pi l,\pi+\sigma^2)\in M_j\cr
&\phi(\sigma^1,2\pi+\sigma^2)=\alpha^k_N\cdot\phi(\sigma^1,\sigma^2)}}

In computing the traces to yield the one--loop expressions, it is
convenient to parameterise the Klein bottle and M\"obius strip in the
region $0\leq\sigma^1\leq4\pi l,0\leq\sigma^2\leq\pi$ as follows:
\eqn\moreparams{\eqalign{{\rm KB}:
\quad&\phi(\sigma^1,\pi+\sigma^2)=\Omega_m\cdot\phi(4\pi
 l-\sigma^1,\sigma^2)\cr
&\phi(4\pi l,\sigma^2)=\alpha^{m-n}_N\cdot\phi(0,\sigma^2)\cr {\rm
MS}:\quad&\phi(0,\sigma^2)\in M_j\cr&\phi(4\pi l,\sigma^2)\in M_j\cr
&\phi(\sigma^1,\pi+\sigma^2)=\Omega_m\cdot\phi(4\pi
l-\sigma^1,\sigma^2).  }} After the standard rescaling of the
coordinates such that open strings are length $\pi$ while closed
strings are length $2\pi$, the amplitudes are
\eqn\trace{\eqalign{
{\rm KB:}\quad &\Tr_{c,k}\left( \Omega_m(-1)^{F+{\tilde
F}}e^{\pi(L_0+{\bar L}_0)/2l}\right)\cr {\rm MS:}\quad
&\Tr_{o,jj}\left( \Omega_m(-1)^{F}e^{\pi L_0/4l}\right)\cr {\rm
C:}\quad &\Tr_{o,ij}\left( \alpha^k_N(-1)^{F}e^{\pi L_0/l}\right).  }}
(Here `$o$' and `$c$' mean `open' and `closed', respectively.)

The complete one--loop amplitude is
\eqn\oneloop{\int_0^\infty{dt\over t} 
\left\{\Tr_c\left({\bf P}(-1)^{\bf F}
e^{-2\pi t(L_0+{\bar L}_0)}\right)+
\Tr_o\left({\bf P}(-1)^{\bf F}e^{-2\pi tL_0} \right) 
\right\}.} 
The projector $\bf P$ includes the GSO and group projections and $\bf
F$ is the spacetime fermion number. The traces are over transverse
oscillator states and include sums over spacetime momenta. After we
evaluate the traces, the $t\to0$ limit will yield the
divergences. Note also that the loop modulus $t$ is related to the
 cylinder length $l$ as $t= 1/4l,1/8l$ and $1/2l$ for the Klein
bottle, M\"obius strip and cylinder, respectively.

In order to compute the loop amplitudes, we must first decide upon a
consistent convention for the action of $\ZN$ and $\Omega$ on the
various sectors of the theory. We follow  most of the conventions used in
ref.\ericjoe. In addition to the conventions listed
there, we have that the elements $\alpha^k_N$ act as follows on the
bosons and in the Neveu--Schwarz (NS) sector:
\eqn\naturalacts{\eqalign{\alpha^k_N:\left\{\eqalign{z_1&
=X^6+iX^7\rightarrow e^{2\pi ik\over N}z_1,\cr
z_2&=X^8+iX^9\rightarrow e^{-{2\pi ik\over N}}z_2}, \right.}} and
it acts in the Ramond (R) sector as \eqn\actsii{\alpha^k_N=e^{{2\pi
ik\over N}(J_{67}-J_{89})}.}  As a consequence of this latter
convention (which has an extra relative minus sign relative to the
analogous operator in ref.\ericjoe), notice for example that
$\alpha^k_N$~gives~$4\!\cos^2\! {\pi k\over N}$ when evaluated on the
R ground states while $(-)^F\alpha^k_N$~gives~$4\!\sin^2\! {\pi k\over
N}$.


\subsec{Loop amplitudes}
Explicitly, we compute the following amplitudes, generalising\ericjoe:
\eqn\compute{\eqalign{&{\rm KB}:\quad \Tr^{U+T}_{NSNS+RR}\left\{ 
{\Omega\over2}\sum_{k=0}^{N-1}
{\alpha^k_N\over N}\cdot{1+(-1)^F\over2}\cdot
e^{-2\pi t(L_0+{\bar L_0})}\right\}    \cr
&{\rm MS}:\quad
\Tr_{NS-R}^{99+55}\left\{{\Omega\over2}\sum_{k=0}^{N-1}{\alpha^k_N\over
N}\cdot{1+(-1)^F\over2}\cdot e^{-2\pi tL_0}\right\}\cr 
&{\rm C}:\quad
\Tr_{NS-R}^{99+55+95+59}\left\{{1\over2}\sum_{k=0}^{N-1}{\alpha^k_N\over
N}\cdot{1+(-1)^F\over2}\cdot
e^{-2\pi tL_0}\right\} ,}} where $U(T)$ refers to
the untwisted (twisted) sector of the closed string. As $\Omega$
forces the left and right moving sector to be identical, there is no
need to include $\half(1+(-1)^{\bar F})$ in the trace in the Klein
bottle. The open string traces include a sum over Chan--Paton factors.

Before listing the results of the careful computations, which yield
the one--loop amplitudes, let us introduce some of the characters and 
notation which will appear.

Playing a central role in organising the amplitudes will be 
 Jacobi's $\vartheta$--functions:
\eqn\thetas
{\eqalign{\vartheta_1(z|t) 
&= 2q^{1/4} \sin \pi z \prod_{n=1}^{\infty} (1-q^{2n})
\prod_{n=1}^{\infty} (1-q^{2n}e^{2\pi iz})\prod_{n=1}^{\infty} 
(1-q^{2n}e^{-2\pi iz})\cr
\vartheta_2(z|t) &= 2 q^{1/4} \cos \pi z\prod_{n=1}^{\infty}
 (1-q^{2n})\prod_{n=1}^{\infty} (1+q^{2n}e^{2\pi iz})
\prod_{n=1}^{\infty} (1+q^{2n}e^{-2\pi iz})\cr 
\vartheta_3(z|t) &=\prod_{n=1}^{\infty} (1-q^{2n})\prod_{n=1}^{\infty} 
(1+q^{2n-1}e^{2\pi iz})\prod_{n=1}^{\infty} (1+q^{2n-1}e^{-2\pi iz})\cr
\vartheta_4(z|t) &=\prod_{n=1}^{\infty} (1-q^{2n})\prod_{n=1}^{\infty}
 (1-q^{2n-1}e^{2\pi iz})\prod_{n=1}^{\infty} (1-q^{2n-1}e^{-2\pi iz})
}} where $q=e^{-\pi t}$.  We will need their asymptotics at $t\to 0$.
The asymptotics as $t\to \infty$ are straightforward.  The asymptotics
as $t\to 0$ are obtained from the modular transformations ($\tau=it$)
\eqn\modthetas{\eqalign{&\vartheta_1(z|\tau) = 
\tau^{-1/2}e^{3i\pi/4}e^{-i\pi z^2/\tau}
\vartheta_1\left({z\over\tau}|-{1\over\tau}\right),\quad
\vartheta_3(z|\tau) = \tau^{-1/2}e^{i\pi/4}
e^{-i\pi z^2/\tau}\vartheta_3\left({z\over\tau}|-{1\over\tau}\right),\cr
&\vartheta_2(z|\tau) = \tau^{-1/2}e^{i\pi/4}
e^{-i\pi z^2/\tau}\vartheta_4\left({z\over\tau}|-{1\over\tau}\right),\quad
\vartheta_4(z|\tau) = \tau^{-1/2}e^{i\pi/4}
e^{-i\pi z^2/\tau}\vartheta_2\left({z\over\tau}|-{1\over\tau}\right).}}
The $f$--functions familiar from ten dimensional string theory 
are a special case of the functions \thetas\ (here, a prime denotes 
$\partial/\partial z$):
\eqn\fs{\eqalign{
f_1(q) &= q^{1/12} \prod_{n=1}^{\infty} (1-q^{2n})
 = (2\pi)^{-1/3}\vartheta_1 ' (0|t)^{1/3}
\cr
f_2(q) &= \sqrt{2} q^{1/12} \prod_{n=1}^{\infty} (1+q^{2n})
 = (2\pi)^{1/6}\vartheta_2 ' (0|t)^{1/2} \vartheta_1 ' (0|t)^{-1/6}
\cr
f_3(q) &= q^{-1/24} \prod_{n=1}^{\infty} (1+q^{2n-1})
 = (2\pi)^{1/6}\vartheta_3 ' (0|t)^{1/2} \vartheta_1 ' (0|t)^{-1/6}
\cr
f_4(q) &= q^{-1/24} \prod_{n=1}^{\infty} (1-q^{2n-1}) =
 (2\pi)^{1/6}\vartheta_4 ' (0|t)^{1/2} \vartheta_1 ' (0|t)^{-1/6} }}
Their asymptotics at $t\to 0$ follow from their modular
transformations and together with their $t\to \infty$ behaviour:
\eqn\modfs
{ f_{1}(e^{-{\pi}/{s}}) = \sqrt{s}\,f_{1}(e^{-\pi s}),\quad
f_{3}(e^{-{\pi}/{s}}) = f_{3}(e^{-\pi s}),\quad
f_{2}(e^{-{\pi}/{s}}) = f_{4}(e^{-\pi s}).}

The familiar {\it ``aequatio identico satis abstrusa''}
\eqn\abstruce{f_3(q)^8-f_4(q)^8-f_2^8(q)=0,}
follows from the more general identities
\eqn\moreabstruce{\eqalign{\vartheta^2_3(0|t)\vartheta^2_3(z|t)- 
\vartheta^2_4(0|t)\vartheta^2_4(z|t)-
\vartheta^2_2(0|t)\vartheta^2_2(z|t)&=0\cr
\vartheta^2_3(0|t)\vartheta^2_2(z|t)- 
\vartheta^2_4(0|t)\vartheta^2_1(z|t)-
\vartheta^2_2(0|t)\vartheta^2_3(z|t)&=0\cr
\vartheta^2_3(0|t)\vartheta^2_4(z|t)- 
\vartheta^2_4(0|t)\vartheta^2_3(z|t)-
\vartheta^2_2(0|t)\vartheta^2_1(z|t)&=0,
}}
of which we will make much use in what is to follow.

The appearance of the full Jacobi $\vartheta$--functions is, in
retrospect, perhaps not surprising, as twisting in the open string
loop channel by $\alpha^k_N$ introduces $z=k/N$ into the oscillator
sums. They therefore arise naturally in the cylinder and M\"obius
strip amplitudes listed below.  In the case of the Klein bottle, there
is also a twist in the closed string loop channel by
$\alpha^{n-m}_N$. Such a space twist will in general change the moding
of the fermion and bosons, producing a `spectral flow' between all of
the different sectors. This should manifest itself as another type of
twist of the $\vartheta$--functions.  To write this relationship, we
use the notation
\eqn\rewrite{\vartheta_1=\vartheta[\textstyle{1\atop1}],
\quad\vartheta_2=\vartheta[{0\atop1}],
\quad\vartheta_3=\vartheta[{0\atop0}],\quad\vartheta_4=\vartheta[{1\atop0}],}
in which we can succinctly write\thetabook:
\eqn\relations{\vartheta[\textstyle{\epsilon\atop\epsilon^\prime}]
(z-\zeta t|t)=
e^{i\pi\{-\tau\zeta^2+\zeta\epsilon^\prime+2\zeta z\}} 
\vartheta[{\epsilon-2\zeta\atop\epsilon^\prime}](z|t),}
where $(\epsilon,\epsilon^\prime)\in\{0,1\}$ for the familiar
$\vartheta$--functions.  In evaluating the Klein bottle amplitude,
the relations \relations\ are used to rewrite twisted expressions in terms of
$\vartheta$--functions.

For the twisted 99 cylinders the one--loop amplitudes are ($z{=}k/N$):
\eqn\Camplitudes{\eqalign{{V_{6}\over2^3N}\sum_{k=1}^{N-1}
{(\Tr(\gamma_{k,9}))^2\over(4\sin^2\pi z)^2}
&\int^\infty_0{dt\over t}(8\pi^2\alpha^\prime t)^{-3}\,\,
4\sin^2\pi z \,\,f_1^{-6}(t)\vartheta^{-2}_1(z|t)\,\,\times\cr
&\left\{\vartheta^2_3(0|t)\vartheta^2_3(z|t)- 
\vartheta^2_4(0|t)\vartheta^2_4(z|t)-
\vartheta^2_2(0|t)\vartheta^2_2(z|t)\right\}
,}} while for the twisted 55 cylinders they are: 
\eqn\Camplitudes{\eqalign{{V_{6}\over2^3N}\sum_{k=1}^{N-1}
{(\Tr(\gamma_{k,5}))^2}
&\int^\infty_0{dt\over t}(8\pi^2\alpha^\prime t)^{-3}\,\,
4\sin^2\pi z \,\,f_1^{-6}(t)\vartheta^{-2}_1(z|t)\,\,\times\cr
&\left\{\vartheta^2_3(0|t)\vartheta^2_3(z|t)- 
\vartheta^2_4(0|t)\vartheta^2_4(z|t)-
\vartheta^2_2(0|t)\vartheta^2_2(z|t)\right\}
.}} 
The 95 cylinders give:
\eqn\Camplitudes{\eqalign{2{V_{6}\over2^3N}\sum_{k=1}^{N-1}
\Tr(\gamma_{k,9})\Tr(\gamma_{k,5})&\int^\infty_0{dt\over t}
(8\pi^2\alpha^\prime t)^{-3}\,\,f_1^{-6}(t)\vartheta^{-2}_4(z|t)\,\,\times\cr
&\left\{\vartheta^2_3(0|t)\vartheta^2_2(z|t)- 
\vartheta^2_4(0|t)\vartheta^2_1(z|t)-
\vartheta^2_2(0|t)\vartheta^2_3(z|t)\right\}
.}}
The twisted M\"obius strip amplitudes are, for the D5--branes ($z{=}m/N$):
\eqn\MSamplitudes{\eqalign{-{V_{6}\over2^3N}\sum_{m=1}^{N-1}
\Tr(\gamma^{-1}_{\Omega_m,5}\gamma^T_{\Omega_m,5})&\int^\infty_0{dt\over t}
(8\pi^2\alpha^\prime t)^{-3}\,\,
4\cos^2\pi z \,\,f_1^{-6}(iq)\vartheta^{-2}_2(iq,z)\,\,\times\cr
&\left\{\vartheta^2_3(iq,0)\vartheta^2_4(iq,z)- 
\vartheta^2_4(iq,0)\vartheta^2_3(iq,z)-
\vartheta^2_2(iq,0)\vartheta^2_1(iq,z)\right\},}}
and for the D9--branes:
\eqn\MSamplitudes{\eqalign{-{V_{6}\over2^3N}\sum_{m=1}^{N-1}
{\Tr(\gamma^{-1}_{\Omega_m,9}\gamma^T_{\Omega_m,9})
\over(4\sin^2\pi z)^2}&\int^\infty_0{dt\over t}
(8\pi^2\alpha^\prime t)^{-3}\,\,
4\sin^2\pi z \,\,f_1^{-6}(iq)\vartheta^{-2}_1(iq,z)\,\,\times\cr
&\left\{\vartheta^2_3(iq,0)\vartheta^2_3(iq,z)- 
\vartheta^2_4(iq,0)\vartheta^2_4(iq,z)-
\vartheta^2_2(iq,0)\vartheta^2_2(iq,z)\right\}.}}
Finally, the Klein bottle gives ($t^+{=}t+\xi t,\, t^-{=}t-\xi t$):
\eqn\KBamplitudes{\eqalign{{V_{6}\over2^3N}&\sum_{m,n=1}^{N-1}
{1\over(4\sin^2\pi z)^2}\int^\infty_0{dt\over t}
(4\pi^2\alpha^\prime t)^{-3}\,\,
4\sin^22\pi (z-\zeta t) \,\,f_1^{-6}(2t)\vartheta^{-1}_1(z|2t^-)
\vartheta^{-1}_1(z|2t^+)\,\,\times\cr
&\left\{-\vartheta^2_4(0|2t)\vartheta_4(z|2t^-)
\vartheta_4(z|2t^+)+ 
\vartheta^2_3(0|2t)\vartheta_3(z|2t^-)
\vartheta_3(z|2t^+)\right.\cr&\hskip5.5cm-
\left.\vartheta^2_2(0|2t)\vartheta_2(z|2t^-)
\vartheta_2(z|2t^+)\right\}.}}
In the Klein bottle amplitudes, we have the twist $\zeta{=}(m-n)/N$ in
the closed string channel, resulting in a zero point energy shift for
the bosons and fermions which contribute.  $V_6$ is the regularised six
dimensional spacetime volume.

The factor of $(4\sin^2\pi z)^{-2}$ is a non--trivial contribution
from  evaluating the trace of the operator ${\cal O}$ in the $z^1$
and $z^2$ complex planes in the NN sector. The operator ${\cal O}$ is the
rotation
\eqn\rot{{\cal O}:\quad z^{1,2}\to e^{\pm{2\pi ik\over N}}z^{1,2}.} We have
\eqn\traces{\Tr[e^{2\pi ik\over N}]=\int dz^1 dz^2<\!z^1,z^2| {\cal O} 
|{z^1}^\prime,{z^2}^\prime\!>=\left(4\sin^2{\pi k\over
N}\right)^{-2},} where we have used the basis 
\eqn\basis{<\!z^1,z^2|{z^1}^\prime,{z^2}^\prime\!>=
{1\over V_{T^4}}\delta(z^1-{z^1}^\prime)\delta(z^2-{z^2}^\prime).}

Supersymmetry is manifest here, as due to the identities
\moreabstruce\ each of these amplitudes vanishes identically. However,
we wish to extract the tadpoles for closed string massless NS-NS fields
from these amplitudes, and we do so by identifying the contribution of
this sector from each of these amplitudes.

\subsec{Factorisation and Tadpoles}
The next step is to extract the asymptotics as $t\to 0$ of the
amplitudes, relating this limit to the $l\to\infty$ limit for each
surface, (using the relation between $l$ and $t$ for each surface
given earlier). Here, the asymptotic behaviour of the $\vartheta$--
and $f$--functions given in equations \modthetas\ and \modfs\ are
used.  This extracts the (divergent) contribution of the massless
closed string R-R fields, which we shall list below.  In what follows,
we shall neglect the overall factors of $1/N$ and powers of 2 which
accompany all of the amplitudes.

First, we list the tadpoles for the untwisted R-R  potentials. For
the 10--form we have the following expression (proportional to
$(1-1)v_{6}v_4\int_0^\infty dl$):
\eqn\tadsone{\Tr(\gamma^{\phantom{-1}}_{0,9})^2 - 64\Tr(\gamma_{\Omega,9}^{-1}
\gamma_{\Omega,9}^T) + 32^2,}
corresponding to the diagrams:

\hskip1.0cm\epsfxsize=3.5in\epsfbox{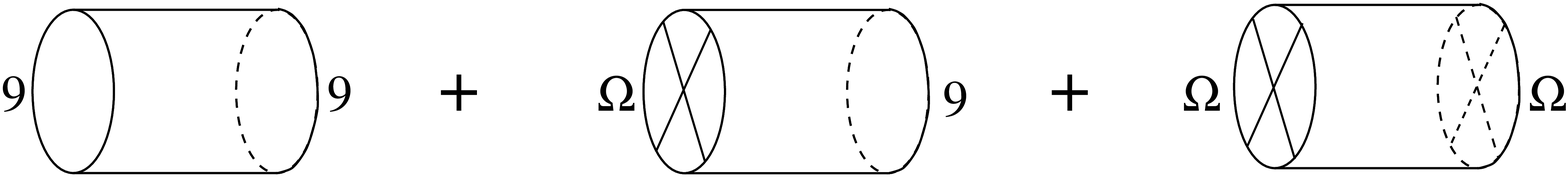}

Here, $v_D=V_D(4\pi^2\alpha^\prime)^{-D/2}$, where $V_D$ is a
regularised $D$ dimensional volume. The limit where we focus upon the
neighbourhood of one ALE point is equivalent to taking the
non--compact limit $v_4\to\infty$ while staying in a frame where our
regulated volume $v_{10}=v_6v_4$ is finite.

For the 6--form we have (proportional to 
$(1-1){v_6\over v_4}\int_0^\infty dl$):
\eqn\tadstwo{\Tr(\gamma^{\phantom{-1}}_{0,5})^2 - 64
\Tr(\gamma_{\Omega_{N\over2},5}^{-1}
\gamma_{\Omega_{N\over2},5}^T) + 32^2,}
which arise from the diagrams:

\hskip1.0cm\epsfxsize=3.5in\epsfbox{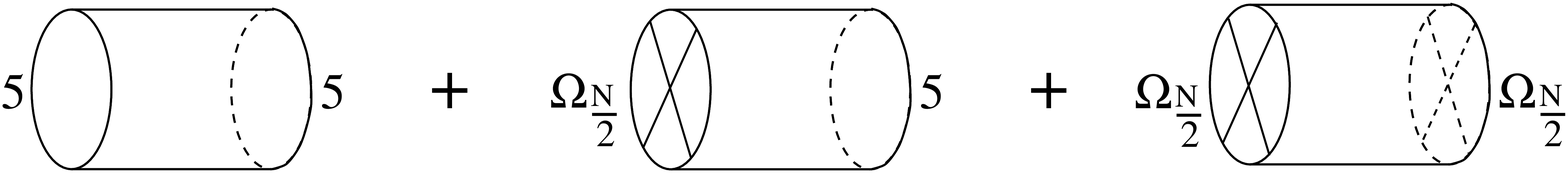}

In the non--compact limit we are considering here, this last
contribution does not survive, as it is proportional to $v_6/v_4$.
The fact that it vanishes is consistent with the fact that if space is
not compact, there is no restriction from charge conservation on the
number of D5--branes which may be present: The analogue of Gauss' Law
for the 6--form potential's field strength does not apply, as the flux
lines can stretch to infinity. In the compact case, they must begin
and end all within the compact volume. So this equation will be
relevant only when we return to the study of global 6--form charge
cancelation in the compact $K3$ examples.

Notice also in this case 
that the last two diagrams obviously vanish in the case
when $N$ is odd. An immediate consequence of this is that ${\bf Z}_3$
fixed points have no untwisted 6--form charge. Their presence alone
will not require dynamical fivebranes.

The twisted sector tadpoles are (proportional to $(1-1)v_6\int_0^\infty dl$):
\eqn\tadpoles{\eqalign{&\sum_{k=1\atop k\neq {N\over2}}^{N-1}
\left[ {1\over 4\sin^2\pkm}  \Tr(\gamma^{\phantom{-1}}_{k,9})^2 - 
2 \Tr(\gamma^{\phantom{-1}}_{k,9})\Tr(\gamma^{\phantom{-1}}_{k,5})+ 
4\sin^2\pkm\Tr(\gamma^{\phantom{-1}}_{k,5})^2 \right]\cr
-16\!\!&\sum_{k=1\atop k\neq{{N\over2}}}^{N-1}\left[4\cos^2{\pkm} 
\Tr(\gamma_{\Omega_k,5}^{-1}\gamma_{\Omega_k,5}^{T})+ 
{1\over 4\sin^2\pkm} 
\Tr(\gamma_{\Omega_k,9}^{-1}\gamma_{\Omega_k,9}^{T})  \right]\cr
+64\!\!&\sum_{{k=1\atop k\neq{{N\over2}}}}^{N-1}
 \left[{\cos^2\pkm\over\sin^2\pkm}-
\delta_{N\,{\rm mod}\, 2,0}\right].
}}
These tadpoles correspond to the following diagrams:
\vskip0.5cm
\hskip0.7cm\epsfxsize=3.5in\epsfbox{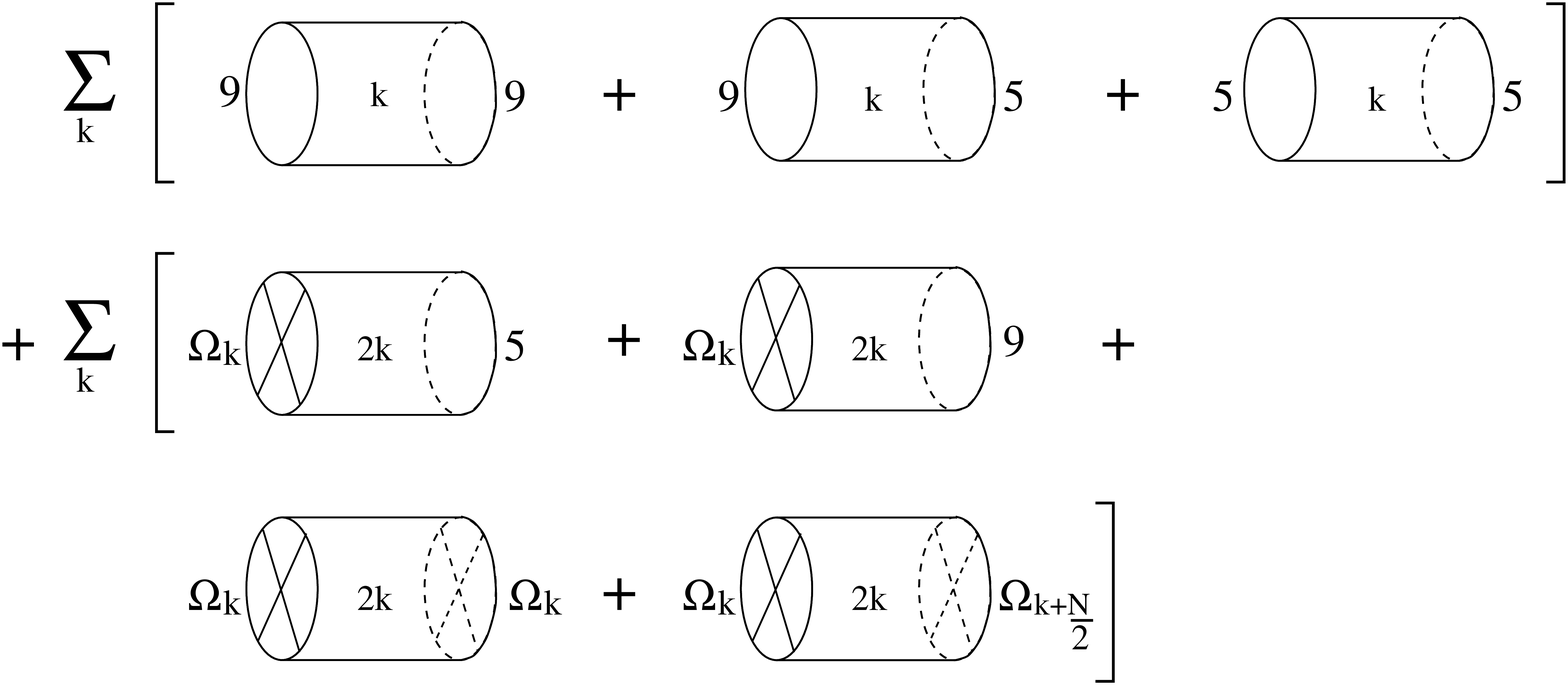}
\vskip0.4cm

Notice that since $\alpha^k_N$ and $\alpha^{k+N/2}_N$ both square to
the same element, $\alpha^{2k}_N$, we can make opposite phase choices
in the composition algebra of the $\gamma^{\phantom{T}}_{\Omega_k}$ matrices:
\eqn\oppositenine{\eqalign{&\Tr[\gamma_{\Omega_k,9}^{-1}
\gamma_{\Omega_k,9}^{T}]=\Tr[\gamma_{2k,9}^{\phantom{-1}}]\cr
&\Tr[\gamma_{\Omega_{k+{N\over2}},9}^{-1}\gamma_{\Omega_{k+{N\over2}},9}^{T}]
=-\Tr[\gamma_{2k,9}^{\phantom{-1}}]
}}  for D9--branes and  
\eqn\oppositefive{\eqalign{&\Tr[\gamma_{\Omega_k,5}^{-1}
\gamma_{\Omega_k,5}^{T}]=-\Tr[\gamma_{2k,5}^{\phantom{-1}}]\cr
&\Tr[\gamma_{\Omega_{k+{N\over2}},5}^{-1}
\gamma_{\Omega_{k+{N\over2}},5}^{T}]=\Tr[\gamma_{2k,5}^{\phantom{-1}}]
}} for D5--branes. This is more than an aesthetic choice, as the first
line of each of these conditions is simply the crucial result derived
in ref.\ericjoe\ that $\Omega^2$=1 in the 99 sector, but $-1$ in the
55 sector. The second line in each is the statement that
$\gamma^2_{N/2}=-1$ in each sector.

With \oppositenine\ and \oppositefive, the expression \tadpoles\ can
be factorised, for even $N$:
\eqn\eventadpoles{\eqalign{&\sum_{k=1}^{{N\over2}}{1\over 4\sin^2\opkm}
\left[\Tr(\gamma^{\phantom{-1}}_{2k-1,9})- 4\sin^2\opkm 
\Tr(\gamma^{\phantom{-1}}_{2k-1,5}) \right]^2\cr
&\sum_{k=1}^{{N\over2}}{1\over 4\sin^2\epkm}
\left[\Tr(\gamma^{\phantom{-1}}_{2k,9})- 4\sin^2\epkm 
\Tr(\gamma^{\phantom{-1}}_{2k,5})-32\cos\epkm \right]^2
}}
and for odd $N$:
\eqn\oddtadpoles{\eqalign{\sum_{k=1}^{M-1}{1\over 4\sin^2\epkm}
\left[\Tr(\gamma^{\phantom{-1}}_{2k,9})- 4\sin^2\epkm 
\Tr(\gamma^{\phantom{-1}}_{2k,5})-32\cos^2\pkm \right]^2
}}

Having extracted the divergences and factorised them, revealing the
tadpole equations (which may be also interpreted as charge
cancelation equations, as discussed earlier) we are ready to find
ways of solving these equations for the various orientifold groups.

\newsec{$K3$ Orientifolds}

\subsec{The  Orientifold Models  and T--Duality.}
Compact manifolds which can be constructed as $T^4/\ZN$ (as described
in section~2) exist only for $N=2,3,4$ and $6$. From the discussion in
section~3, we can therefore construct orientifolds of type $A$ for all
these $N$, but of type $B$ only for $N=2,4$ and $6$.  We list below
explicitly the orientifold groups:
\eqn\theorientifolds{\eqalign{&\Z_2^A=\{1,\alpha^1_2,\Omega,
\Omega \alpha^1_2\},\quad\Z_2^B=\{1,\Omega \alpha_2^1\},\cr
&\Z_3^A=\{1,\alpha_3^1,\alpha_3^2,\Omega, \Omega\alpha_3^1,
\Omega\alpha_3^2 \},\cr &\Z_4^A=\{1, \alpha^1_4,\alpha_4^2,\alpha^3_4, \Omega,
\Omega\alpha^1_4,\Omega\alpha_4^2,\Omega\alpha^3_4,\},
\quad\Z_4^B=\{1,\alpha_4^2,
 \Omega\alpha^1_4,\Omega\alpha^3_4,\}\cr &\Z_6^A=\{1,
\alpha^1_6,\ldots,\alpha^5_6,
\Omega,\Omega\alpha^1_6,\ldots,\Omega\alpha^5_6
\},\cr&\Z_6^B=\{1,\alpha_6^2,\alpha^4_6,\Omega\alpha^1_6,
\Omega\alpha_6^3,\Omega\alpha^5_6\},
}} where $\alpha_N^{N\over2}\equiv R$. 
\def\gamp{\gamma^{\phantom{-1}}}
In equation \tadsone\ for the untwisted 10--form potential,
$\Tr(\gamp_{0,9})=n_9$, the number of D9--branes. All of the
orientifold groups of type $A$ contain the element $\Omega$, and
therefore there will be an equation of the form \tadsone, telling us
that there are 32 D9--branes. Similarly, all type $A$ models except
$\Z_3^A$ will contain 32 D5--branes also, as the presence of an
element $\Omega R$ means that there will be an equation of the form
\tadstwo.

In contrast, the models of type $B$ all lack the element $\Omega$ and
therefore have only the first term of equation \tadsone. Therefore the
number of D9--branes in these models is zero. All type $B$ models
except $\Z_4^B$ have the element $\Omega R$, and therefore have 32
D5--branes.  So $\Z_4^B$ has the distinction of having no open string
sectors at all: It is a consistent unoriented closed string
theory\foot{There are other examples in the literature\edvafa,
constructed as an orientifold together with a translation. Later in
the paper, we will find a surprise: When we compute the spectrum of
our closed string model, it is the same as that of the closed string
sector of the orientifold model of ref.\atish.}.

As already discussed in section~3, T--duality in the $(6,7,8,9)$
directions exchanges the elements $\Omega$ and $\Omega R$. This also
exchanges D9--branes with D5--branes. So one might imagine that there
are some T--duality relations amongst the models, which is of course
true: Models $\Z_2^A$, $\Z_4^A$, $\Z_6^A$ and $\Z^B_4$ are self
T$_{6789}$--dual. They contain the same numbers of D--branes of each
type. Meanwhile $\Z_3^A$, which has only D9--branes, is dual to
$Z_6^B$ which has only D5--branes. $\Z_2^B$, which has only $\Omega R$
as a non--trivial element of its orientifold group, is dual to
ordinary Type~I string theory (which we may denote as $\Z_1^A$), whose
orientifold group has only $\Omega$ as its non--trivial element.

To summarise, we have the following picture for the T--duality
relationships among the models:
\vskip0.5cm
\hskip3.5cm\epsfxsize=2.0in\epsfbox{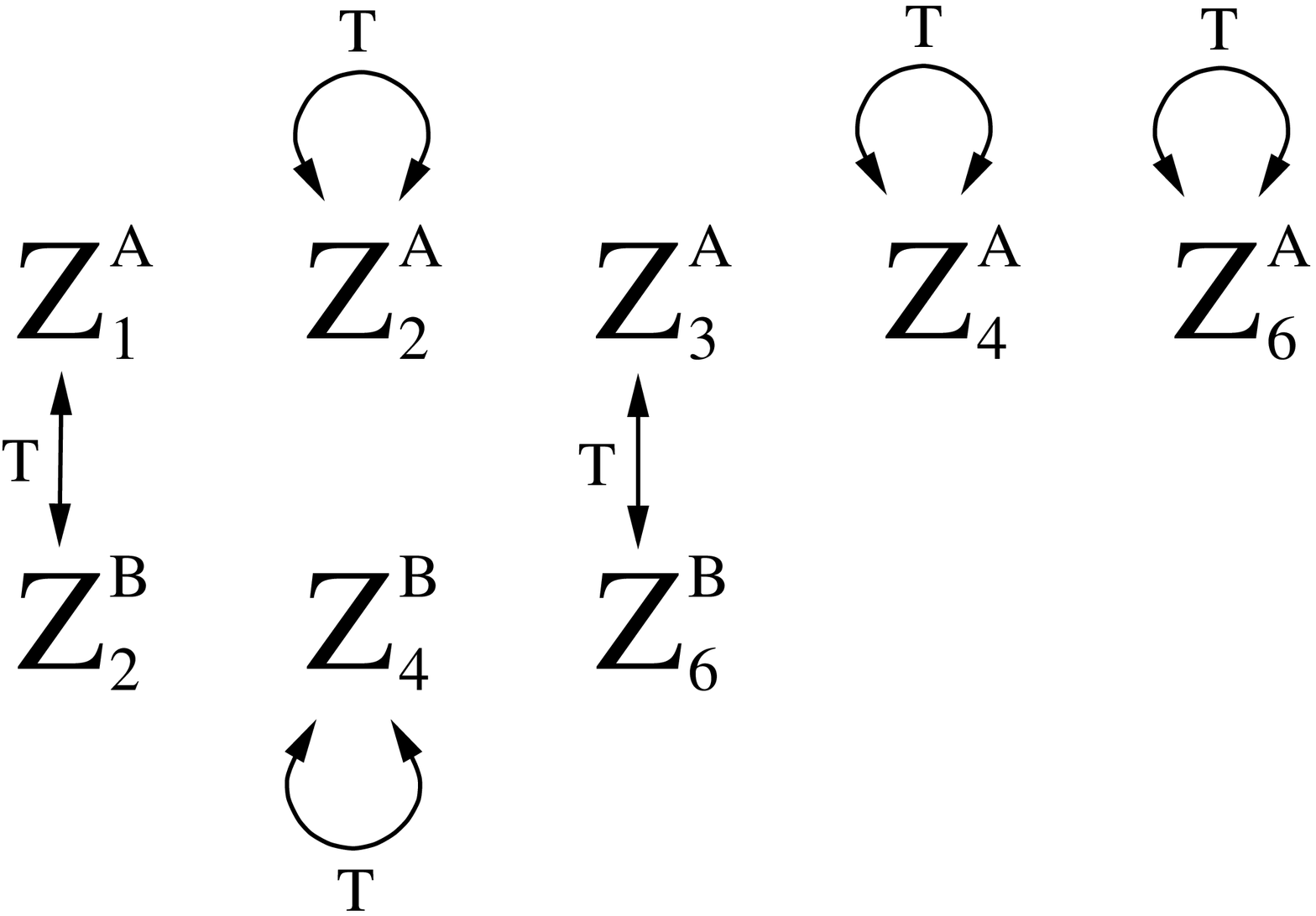}
\vskip0.4cm

Having established the models we wish to consider, let us revisit the
tadpole equations and study them some more.

At this stage, the notation with which we compactly carried out the
tadpole calculation to this point is a now more of a hindrance than an
aid to clarity.  Much is to be gained by simply writing out the
tadpole equations explicitly in each case.

For purposes of comparison, we start with the already
computed\ericjoe\ $\Z_2^A$ case, for which there is one twisted
tadpole equation (recall $\alpha^1_2\equiv R$):
\eqn\ztwo{\Tr[\gamp_{1,9}]-4\Tr[\gamp_{1,5}]=0.}
The basic solution was found to be 
\eqn\ztwotwo{\eqalign{\gamp_{\Omega,9}&=\gamp_{\Omega R,5}=I_{32}\cr
\gamp_{R,9}&=\gamp_{\Omega,5}=\left(\matrix{0&I_{16}\cr-I_{16}&0}\right).
}}

For the $\Z_3^A$ case we have 
\eqn\zthree{\eqalign{&\Tr[\gamp_{1,9}]-3\Tr[\gamp_{1,5}]=8\cr
&\Tr[\gamp_{2,9}]-3\Tr[\gamp_{2,5}]=8,}} and since we have already
learned that the number of D5--branes is zero in this case, we have
solution $\gamma_{5}=0$ for all orientifold elements in the D5--brane
sector, and we can write 
\eqn\zthreetwo{\eqalign{\gamp_{\Omega,9}&=\left(\matrix{0&1&0&0\cr
1&0&0&0\cr0&0&1&0\cr0&0&0&1}\right)\otimes I_{8}, \cr
\gamp_{1,9}&={\rm diag}\{ e^{2\pi i\over3}\,(8\,\,{\rm times }), 
e^{-{2\pi i\over3}}\,(8\,\,{\rm times }),1\,(16\,\,{\rm times })\},}}
from which it is trivially verified that \zthree\ is satisfied. 

Notice that $\gamma_\Omega$ acts by exchanging the roots and their
complex conjugates: $e^{{2\pi i\over3}}\leftrightarrow e^{-{2\pi
i\over3}} $. This will be the case in all of the later models, and so
we will no longer list it explicitly in the later solutions. Note also
that in the other type $A$ orientifolds,
$\gamma_{\Omega,9}=\gamma_{\Omega R,5}$, and $\gamma_{\Omega
R,9}=\gamma_{\Omega,5}$. It can also be shown that we can choose a
phase such that we can always write $\gamma_{1,9}=e^{2\pi im\over
N}\gamma_{1,5}$, for $m$ any odd integer. That we can find such a
simple relationship between the $\gamma$ matrices in the D5-- and
D9--brane sectors is a manifestation of T$_{6789}$ duality.

For the $\Z_4^A$ case we have:
\eqn\zfour{\eqalign{&\Tr[\gamp_{1,9}]-2\Tr[\gamp_{1,5}]=0\cr
&\Tr[\gamp_{2,9}]-4\Tr[\gamp_{2,5}]=0\cr
&\Tr[\gamp_{3,9}]-2\Tr[\gamp_{3,5}]=0.}}
Note that the middle case correctly reproduces the $\Z_2^A$ example, and
therefore the $\Z_2^A$ example appears as a substructure. This will be
true for $\Z_6^A$ also.

The  solution is ($\alpha^2_4\equiv R$)
\eqn\zfourtwo{\eqalign{\gamp_{1,9}=
&{\rm diag}\{e^{\pi i\over4}\,(8\,\,{\rm times }), 
e^{-{\pi i\over4}}\,(8\,\,{\rm times }),
e^{3\pi i\over4}\,(8\,\,{\rm times }),
e^{-{3\pi i\over4}}\,(8\,\,{\rm times }) \}
}}

For the $\Z_6$ case we have:
\eqn\zsix{\eqalign{&\Tr[\gamp_{1,9}]-\Tr[\gamp_{1,5}]=0\cr
&\Tr[\gamp_{2,9}]-3\Tr[\gamp_{2,5}]=16\cr
&\Tr[\gamp_{3,9}]-4\Tr[\gamp_{3,5}]=0\cr
&\Tr[\gamp_{4,9}]-3\Tr[\gamp_{4,5}]=-16\cr
&\Tr[\gamp_{5,9}]-\Tr[\gamp_{5,5}]=0
,}}
for which we have ($\alpha^2_4\equiv R$)
\eqn\zsixtwo{\eqalign{&\Tr[\gamp_{1,9}]=\Tr[\gamp_{1,5}]=0,\,\,
\Tr[\gamp_{3,9}]=\Tr[\gamp_{3,5}]=0,\,\,
\Tr[\gamp_{5,9}]=\Tr[\gamp_{5,5}]=0,\cr
&\Tr[\gamp_{2,9}]=\Tr[\gamp_{2,5}]=-8,\,\,
\Tr[\gamp_{4,9}]=\Tr[\gamp_{4,5}]=8,\cr
&\gamp_{1,9}= {\rm diag}\{e^{\pi i\over6}\,(4\,\,{\rm times }), 
e^{-{\pi i\over6}}\,(4\,\,{\rm times }),\cr&\hskip3cm
e^{5\pi i\over6}\,(4\,\,{\rm
times }), e^{-{5\pi i\over6}}\,(4\,\,{\rm times }), -i\,(8\,\,{\rm
times }),i\,(8\,\,{\rm times })\} .}}  Note here that
\eqn\note{\gamp_{1,9}\equiv{\rm diag}\{e^{2\pi i\over3}\,(4\,\,{\rm times }), 
e^{-{2\pi i\over3}}\,(4\,\,{\rm times }),1\,(8\,\,{\rm times
})\}\otimes {\rm diag}\{-i,i\},} which shows a $\Z_3\times\Z_2$
structure, using the solutions previously obtained for the $\Z_2^A$
and $\Z_3^A$ models.

Also notice that in all cases above, the coefficient of the
$\gamma_{k,5}$ trace is the square root of the number of fixed points
invariant under $\alpha^k_N$. Also interesting is that (generalising
the $\Z_2$ case,) the same choice made for D9--branes can be made for
D5--branes, up to a phase.

The tadpoles for the case $\Z_6^B$ will turn out to be isomorphic to
those listed above for $\Z_3^A$, while there are no tadpoles to list
for the $\Z_4^B$ model, as there are no D--branes required.

Let us now turn to the closed string spectra.

\subsec{Closed String Spectra}
The right moving untwisted sector has the massless states:
\eqn\spectri{\matrix{{\rm Sector}&{\rm State}&\alpha^k_N&SO(4)
\,\,{\rm rep.}\cr\cr
{\rm NS}:&\psi^\mu_{-1/2}|0\!>&1&\bf{(2,2)}\cr
       &\psi^{1\pm}_{-1/2}|0\!>&e^{\pm{2\pi ik\over N}}&2\bf{(1,1)}\cr
       &\psi^{2\pm}_{-1/2}|0\!>&e^{\mp{2\pi ik\over N}}&2\bf{(1,1)}\cr
       {\rm R}:&|s_1s_2s_3s_4\!>&&&\cr
       &s_1=+s_2,\,s_3=+s_4&1&2\bf{(2,1)}\cr
       &s_1=-s_2,\,s_3=-s_4&e^{\pm{2\pi ik\over N}}&2\bf{(1,2)}\cr }}
       while the right moving sector twisted by ${m\over N}\neq
       {1\over2}$ has:
\eqn\spectrii{\matrix{{\rm Sector}&{\rm State}&\alpha^k_N&SO(4)
\,\,{\rm rep.}\cr\cr
{\rm NS}:&\psi^{1+}_{-1/2+m/N}|0\!>&e^{{2\pi ik\over N}(1-2{m\over N})}
         &\bf{(1,1)}\cr 
         &\psi^{2+}_{-1/2+m/N}|0\!>& e^{{2\pi ik\over N}(1-2{m\over N})} 
         &\bf{(1,1)}\cr 
         {\rm R}:&|s_1s_2\!>,\,\,s_1=-s_2&
          e^{{2\pi ik\over N}(1-2{m\over N})}
 &\bf{(1,2)}.  }}

The exception to this situation is when we have a ${m\over
N}{=}{1\over2}$ twist:
\eqn\spectriii{\matrix{{\rm Sector}&{\rm State}&\alpha^k_N&SO(4)
\,\,{\rm rep.}\cr\cr
{\rm NS}:&|s_3s_4\!>,\,\,s_3=+s_4&1&2\bf{(1,1)}\cr {\rm
R}:&|s_1s_2\!>,\,\,s_1=-s_2&1&\bf{(1,2)}.  }}

We have imposed the GSO projection, and decomposed the little group of
the spacetime Lorentz group as $SO(4)=SU(2){\times}SU(2)$. We form the
spectrum for orientifold group of type $A$ by taking products of states from
the left and right sectors (to give states invariant under
$\alpha^k_N$), symmetrised by the $\Omega$ projection in the NS-NS
sector, while antisymmetrising in the R-R sector.

Finally, we have  from the untwisted closed string sector of the type
$A$ $\ZN$ orientifold ($N{\neq}2$):
\eqn\spectri{\matrix{{\rm Sector}&SO(4)\,\,{\rm rep.}\cr\cr
{\rm NS\,NS}:&\bf{(3,3)}+5\bf{(1,1)}\cr {\rm
R\,R}:&\bf{(3,1)}+{\bf(1,3)}+4\bf{(1,1)}.  }} This is the content of
the ${\cal N}{=}1$ supergravity multiplet in six dimensions, accompanied
by one tensor multiplet and 2 hypermultiplets. In the case of $\Z_2$
it is\ericjoe:
\eqn\spectri{\matrix{{\rm Sector}&SO(4)\,\,{\rm rep.}\cr\cr
{\rm NS\,NS}:&\bf{(3,3)}+11\bf{(1,1)}\cr {\rm
R\,R}:&\bf{(3,1)}+{\bf(1,3)}+6\bf{(1,1)}, }} the $D{=}6$, ${\cal N}{=}1$
supergravity multiplet in six dimensions, accompanied by one tensor
multiplet and 4 hypermultiplets 

The twisted sectors will produce additional multiplets. The bosonic
content of a hypermultiplet is four scalars $4{\bf(1,1)}$, while that
of a tensor multiplet is ${\bf (3,1)}+{\bf(1,1)}$. By combining on the
left and right the sectors twisted by $m\over N$ and $(1-{m\over N})$,
we find that the NS-NS sector produces one hypermultiplet while the
R-R sector produces a tensor multiplet. A sector twisted by $1\over2$
simply produces one hypermultiplet: one quarter coming from the R-R
sector and three quarters from the NS-NS sector.

To evaluate\walton\ the number of hypermultiplets coming from the
twisted sectors or a $K3$ orbifold, we need to recall the structure of
the fixed points and their transformation properties, as reviewed in
section~2.  In the case of $\Z_2^A$, we simply multiply by the number of
$\Z_2$ fixed points, and we find that there are 16 hypermultiplets
from the twisted sectors, giving a total of 20 hypermultiplets when
combined with the four from the untwisted sector.  

For $\Z_3^A$ there
are 9 fixed points, each supplying a hypermultiplet and a tensor
multiplet, (for twists by (${1\over3},{2\over3})$), giving a total of 11
hypermultiplets and 10 tensor multiplets when added to those arising
in the untwisted sector. 

For $\Z_4^A$ the four $\Z_4$ invariant fixed
points give 4 hypermultiplets and four tensor multiplets. They are
also $\Z_2$ fixed points and so supply an additional 4
hypermultiplets. The other 12 $\Z_2$ fixed points form 6 $\Z_4$
invariant pairs, from which arise 6 hypermultiplets. This gives a
total of 16 hypermultiplets and 5 tensor multiplets for the complete
model.  

Finally, for the model $\Z_6^A$, the $\Z_6$ fixed point gives 2
hypermultiplet and 2 tensor multiplet from (${1\over6},{5\over6}$) and
(${1\over3},{2\over3}$) twists. It also gives 6 hypermultiplets from the
$1\over2$ twisted sector. The 4 pairs of $\Z_3$ points give 4 tensor
multiplets and 4 hypermultiplets for (${1\over3},{2\over3}$) twists,
while the 5 $\Z_2$ triplets of fixed points supply 5 hypermultiplets.
This gives 14 hypermultiplets and 7 tensor multiplets in all.

For $\ZN$ orientifolds of type $B$ the situation is as follows. For
closed string states, prior to making the theory unorientable, the
relevant orbifold states to consider are those for the group formed by
the remaining pure $\ZN$ elements in the orientifold group, which is
therefore $\Z_{N\over2}$.  The possible left and right states are
evaluated as before, and then they are projected to the unoriented theory 
invariant under   $\Omega\cdot\alpha^1_N$.

It is thus easy to see that the closed string spectra for $\Z^B_6$ and
$\Z_3^A$ are isomorphic, as are those of $\Z_2^B$ and $\Z_1^A$, (the
latter being simply ten dimensional Type~I string theory: there is no
orbifold to perform for $\Z_2^B$).

There remains only the spectrum of $\Z_4^B$ to compute, which is self
T--dual. The pure orbifold states to consider are those of $\Z_2$.
Tensoring left and right to form the $\Omega_1$ invariant
spectrum, we obtain 12 hypermultiplets and 9 tensor multiplets in total.

In summary, we have (in addition to the usual gravity and tensor  multiplet)
the following spectrum of 
hypermultiplets and  tensor multiplets from the closed string sector for
 each model:

\bigskip
\vbox{
$$\vbox{\offinterlineskip
\hrule height 1.1pt
\halign{&\vrule width 1.1pt#
&\strut\quad#\hfil\quad&
\vrule#
&\strut\quad#\hfil\quad&
\vrule#
&\strut\quad#\hfil\quad&
\vrule width 1.1pt#\cr
height3pt
&\omit&
&\omit&
&\omit&
\cr
&\hfil Model&
&\hfil \vbox{\hbox{\hskip0.7cm Neutral}\vskip3pt\hbox{Hypermultiplets}}&
&\hfil \vbox{\hbox{\hskip0.9cm Extra}\vskip3pt\hbox{Tensor Multiplets}}&
\cr
height3pt
&\omit&
&\omit&
&\omit&
\cr
\noalign{\hrule height 1.1pt}
height3pt
&\omit&
&\omit&
&\omit&
\cr
&\hfil $\Z_2^A$&
&\hfil 20&
&\hfil 0&
\cr
height3pt
&\omit&
&\omit&
&\omit&
\cr
\noalign{\hrule}
height3pt
&\omit&
&\omit&
&\omit&
\cr
&\hfil $\Z_3^A$&
&\hfil 11&
&\hfil 9&
\cr 
height3pt 
&\omit& 
&\omit& 
&\omit&
\cr
\noalign{\hrule }
height3pt 
&\omit& 
&\omit& 
&\omit&
\cr
&\hfil $\Z_4^A$&
&\hfil 16&
&\hfil 4&
\cr
height3pt 
&\omit& 
&\omit& 
&\omit&
\cr
\noalign{\hrule } 
height3pt 
&\omit& 
&\omit& 
&\omit&
\cr
&\hfil $\Z_6^A$&
&\hfil 14&
&\hfil 6&
\cr 
height3pt 
&\omit& 
&\omit& 
&\omit&
\cr
\noalign{\hrule } 
height3pt
&\omit& 
&\omit& 
&\omit&
\cr
&\hfil $\Z_4^B$&
&\hfil 12&
&\hfil 8&
\cr
height3pt 
&\omit& 
&\omit& 
&\omit&
\cr
\noalign{\hrule } 
height3pt 
&\omit& 
&\omit& 
&\omit&
\cr
&\hfil $\Z_6^B$&
&\hfil 11&
&\hfil 9&\cr
}
\hrule height 1.1pt}
$$
}

Of course, the fact that we have obtained, in addition to the usual
supergravity multiplet and tensor multiplet, a total of 20
hypermultiplets plus tensor multiplets (80 scalar fields) in the
different orbifold limits of $K3$ should alert us that we are
computing an invariant property of the manifold: there are 80 moduli
of the $K3$ surface\seiberg. It is a four dimensional manifold and so
they should naturally combine into 20 hypermultiplets if all of these
moduli were available to us. This is what happened in the case of
$\Z_2^A$, as can be seen above. However, in the other other
orientifold examples, some of the moduli scalars combine into tensor
multiplets, leaving us with fewer hypermultiplets in the final model,
presumably corresponding to a reduction in the dimension of the moduli
space of $K3$ deformations available to these models.

We now turn to the open string spectra.

\subsec{Open String Spectra}
Let us study first the 99 open string sector. The massless bosonic spectrum
 arises as follows:
\eqn\spectra{\eqalign{\phantom{boo!}&\matrix{&{\rm state}   &\alpha^k_N=+ 
 &\Omega=+&SO(4)\quad {\rm rep.}\cr\phantom{boo!} &
\phantom{\psi^\mu_{-1/2}|0,ij\!>\lambda_{ij}}& 
\phantom{\lambda=\gamp_{k,9}\lambda\gamma^{-1}_{k,9}}&
\phantom{\lambda=-\gamp_{\Omega,9}\lambda^T\gamma^{-1}_{\Omega,9} }
&\phantom{(2,2)}
}\cr& 99\left\{
\matrix{
\psi^\mu_{-1/2}|0,ij\!>\lambda_{ij}& \lambda=\gamp_{k,9}
\lambda\gamma^{-1}_{k,9}&\lambda=-\gamp_{\Omega,9}\lambda^T
\gamma^{-1}_{\Omega,9} &\bf{(2,2)}\cr
\psi^{1\pm}_{-1/2}|0,ij\!>\lambda_{ij}& \lambda=e^{\pm{2\pi 
k\over N}}\gamp_{k,9}\lambda\gamma^{-1}_{k,9}&\lambda=
-\gamp_{\Omega,9}\lambda^T\gamma^{-1}_{\Omega,9}  &2\bf{(1,1)}\cr
\psi^{2\pm}_{-1/2}|0,ij\!>\lambda_{ij}& \lambda=e^{\mp{2\pi k\over N}}
\gamp_{k,9}\lambda\gamma^{-1}_{k,9}&\lambda=-\gamp_{\Omega,9}\lambda^T
\gamma^{-1}_{\Omega,9}  &2\bf{(1,1)}
}\right.
}}
For the 55 states at a fixed point we have: 
\eqn\spectre{\eqalign{\phantom{boo!}&\matrix{&{\rm state}   &\alpha^k_N=+ 
 &\Omega=+&SO(4)\quad {\rm rep.}\cr\phantom{boo!} &
\phantom{\psi^\mu_{-1/2}|0,ij\!>\lambda_{ij}}& \phantom{\lambda=\gamp_{k,5}
\lambda\gamma^{-1}_{k,5}}&\phantom{\lambda=-\gamp_{\Omega,5}\lambda^T
\gamma^{-1}_{\Omega,5} }&\phantom{(2,2)}
}\cr&55\left\{
\matrix{
\psi^\mu_{-1/2}|0,ij\!>\lambda_{ij}& \lambda=\gamp_{k,5}\lambda
\gamma^{-1}_{k,5}&\lambda=-\gamp_{\Omega,5}\lambda^T\gamma^{-1}_{\Omega,5}
 &\bf{(2,2)}\cr
\psi^{1\pm}_{-1/2}|0,ij\!>\lambda_{ij}& \lambda=e^{\pm{2\pi k\over N}}
\gamp_{k,5}
\lambda\gamma^{-1}_{k,5}&\lambda=\gamp_{\Omega,5}\lambda^T
\gamma^{-1}_{\Omega,5} 
 &2\bf{(1,1)}\cr
\psi^{2\pm}_{-1/2}|0,ij\!>\lambda_{ij}& \lambda=e^{\mp{2\pi
 k\over N}}\gamp_{k,5}
\lambda\gamma^{-1}_{k,5}&\lambda=\gamp_{\Omega,5}\lambda^T
\gamma^{-1}_{\Omega,5} 
 &2\bf{(1,1)}
}\right.}}
For the 55 states away from a fixed point we have: 
\eqn\spectre{\eqalign{\phantom{boo!}&\matrix{&{\rm state}  
 &\Omega=+&SO(4)\quad 
{\rm rep.}\cr\phantom{boo!} &
\phantom{\psi^\mu_{-1/2}|0,ij\!>\lambda_{ij}}&\phantom
{\lambda=-\gamp_{\Omega,5}
\lambda^T\gamma^{-1}_{\Omega,5} }&\phantom{(2,2)}
}\cr&55\left\{
\matrix{
\psi^\mu_{-1/2}|0,ij\!>\lambda_{ij}&\lambda=-\gamp_{\Omega,5}
\lambda^T\gamma^{-1}_{\Omega,5} &\phantom{boo!}\bf{(2,2)}\cr
\psi^{1\pm}_{-1/2}|0,ij\!>\lambda_{ij}&\lambda=\gamp_{\Omega,5}
\lambda^T\gamma^{-1}_{\Omega,5}  &\phantom{boo!}2\bf{(1,1)}\cr
\psi^{2\pm}_{-1/2}|0,ij\!>\lambda_{ij}&\lambda=\gamp_{\Omega,5}
\lambda^T\gamma^{-1}_{\Omega,5}  &\phantom{boo!}2\bf{(1,1)}
}\right.}}

For the 59 states we have at a fixed point: 
\eqn\spectre{\eqalign{\phantom{boo!}&\matrix{&{\rm state}   
&\alpha^k_N=+  &\hskip0.5cm SO(4)\quad {\rm rep.}\cr\phantom{boo!} &
\phantom{|s_3s_4,ij\!>\lambda_{ij},\,\,s_3=-s_4}& 
\phantom{\lambda=\gamp_{k,9}\lambda\gamma^{-1}_{k,9}}&\phantom{(2,2)}
}\cr&59:\quad\quad
\matrix{
|s_3s_4,ij\!>\lambda_{ij},s_3=s_4& \lambda=\gamp_{k,5}
\lambda\gamma^{-1}_{k,9}&\phantom{boo!}\bf{2(1,1)}}}}
and  away from a fixed point: 
\eqn\spectre{\eqalign{\phantom{boo!}&\matrix{&{\rm state}  
  &\hskip0.5cm SO(4)\quad {\rm rep.}\cr\phantom{boo!} &
\phantom{|s_3s_4,ij\!>\lambda_{ij},\,\,s_3=s_4}&\phantom{2(1,1)}
}\cr&59:\quad\quad
\matrix{
|s_3s_4,ij\!>\lambda_{ij},\,\,s_3=s_4&\phantom{boo!}2\bf{(1,1)}}}}

Using the solution presented in section~4 for the $\gamma$ matrices,
we find the following solutions for the open string spectra of the models:

\bigskip

\vbox{
$$\vbox{\offinterlineskip
\hrule height 1.1pt
\halign{&\vrule width 1.1pt#
&\strut\quad#\hfil\quad&
\vrule#
&\strut\quad#\hfil\quad&
\vrule#
&\strut\quad#\hfil\quad&
\vrule width 1.1pt#\cr
height3pt
&\omit&
&\omit&
&\omit&
\cr
&\hfil Model&
&\hfil Gauge Group&
&\hfil Charged Hypermultiplets&
\cr
height3pt
&\omit&
&\omit&
&\omit&
\cr
\noalign{\hrule height 1.1pt}
height3pt
&\omit&
&\omit&
&\omit&
\cr
&\hfil $\Z_2^A$&
& $\eqalign{&99:\quad U(16)\cr &55:\quad U(16)\cr&\phantom{59:\quad}}$&
& $\eqalign{&99:\quad 2\times {\bf 120}\cr&55:\quad 2 \times {\bf 120}\cr
&59:\quad ({\bf 16,16})}$ &
\cr
height3pt
&\omit&
&\omit&
&\omit&
\cr
\noalign{\hrule}
height3pt
&\omit&
&\omit&
&\omit&
\cr
&\hfil $\Z_3^A$&
& $\eqalign{ 
99:\quad U(8)\times SO(16)\cr
}$&
&  $\eqalign{
99:\quad  {\bf (28,1)}; \,\,{\bf (8,16)}\cr
}$&
\cr 
height3pt 
&\omit& 
&\omit& 
&\omit&
\cr
\noalign{\hrule }
height3pt 
&\omit& 
&\omit& 
&\omit&
\cr
&\hfil $\Z_4^A$&
& $\eqalign{ 
&99:\quad U(8)\times U(8)\cr
&55:\quad U(8)\times U(8)\cr
&\phantom{59:\quad}}$&
& $\eqalign{
&99:\quad {\bf (28,1)};\,\,\,{\bf (1,28)};\,\,\,{\bf (8,8)}\cr
&55:\quad {\bf (28,1)};\,\,\,{\bf (1,28)};\,\,\,{\bf (8,8)}\cr
&59:\quad ({\bf 8,1;8,1});\,\,\,({\bf 1,8;1,8})}$ &
\cr
height3pt 
&\omit& 
&\omit& 
&\omit&
\cr
\noalign{\hrule } 
height3pt 
&\omit& 
&\omit& 
&\omit&
\cr
&\hfil $\Z_6^A$&
& $\eqalign{ 
&99:\quad U(4)\times U(4)\times U(8)\cr
&55:\quad U(4)\times U(4)\times U(8)\cr
&\phantom{59:\quad}}$&
& $\eqalign{
99:\quad &{\bf (6,1,1)};\,\,\,{\bf (1,6,1)}\cr
&{\bf (4,1,8)};\,\,\,{\bf (1,4,8)}\cr
55:\quad &{\bf (6,1,1)};\,\,\,{\bf (1,6,1)}\cr
&{\bf (4,1,8)};\,\,\,{\bf (1,4,8)}\cr
59:\quad &{\bf (4,1,1;4,1,1)}\cr
&{\bf (1,4,1;1,4,1)}\cr
&{\bf (1,1,8;1,1,8)}}$ &
\cr 
height3pt 
&\omit& 
&\omit& 
&\omit&
\cr
\noalign{\hrule } 
height3pt
&\omit& 
&\omit& 
&\omit&
\cr
&\hfil $\Z_4^B$&
&\hfil {\rm ---}&
&\hfil {\rm ---}&
\cr
height3pt 
&\omit& 
&\omit& 
&\omit&
\cr
\noalign{\hrule } 
height3pt 
&\omit& 
&\omit& 
&\omit&
\cr
&\hfil $\Z_6^B$&
& $55:\quad U(8)\times SO(16)$&
& $55:\quad {\bf (28,1)};\,\,\,{\bf (8,16)}$ &\cr
}
\hrule height 1.1pt}
$$
}

\newsec{Small Instantons and the Dynamics of Fivebranes}
In the table above, we have given the spectrum for comfigurations with
all of the D5--branes sitting on a single fixed point.  There are
other models corresponding to different configurations.  It is easy to
see that some configurations can be reached from other models in the
same moduli space: For fixed the $A$ type orientifolds, a group of
$2N$ D5--branes can move off a fixed point together, forming a
`dynamical fivebrane'.  The 55 hypermultiplets supply precisely the
structure needed to act as moduli for the process of moving off a
fixed point.

For example, in the case of $\Z_4$, four dynamical five--branes (8
D5--branes each) can to move off the fixed point into the bulk,
breaking the group to a $U(6){\times}U(6)$ factor from the remaining
24 D5--branes on the fixed point and a $SU(2)$ factor from the
fivebrane in bulk. If $m\leq4$ of these fivebranes move off, the group
factor at the fixed point is $U(8-2m)\times U(8-2m)$, and away from it
is $SU(2)^m$ for $m$ separated fivebranes, or $USp(2m)$ if they
coincide. It can also be $U(m)\times U(m)$ if the fivebranes move to
another fixed point.  The spectra for these configurations are
accompanied by hypermultiplets in the obvious analogous
representations from the 55 and 59 sectors. The resulting gauge groups
and hypermultiplets are precisely those obtained from the Higgs
mechanism, whereby some of the gauge fields swallow hypermultiplets
from the 55 sector.

For the case $Z_6$ the dynamics of fivebranes is as follows. The
hypermultiplet structure permits the Higgs mechanism corresponding to
the movement of 2 dynamical hypermultiplets (12 D5--branes each). A
single such object at a fixed point gives
$U(2){\times}U(2){\times}U(2)$ contribution to the gauge group. Away
from a fixed point it is again $SU(2)$.

Notice that in this case there is always a residue of 8 D5--branes
(equivalent to 2/3 fivebrane) which must sit at either all at the
$\Z_6$ fixed point, or distributed amongst the $\Z_3$ points in a
$\Z_6$ and $\Omega$ invariant way\foot{This is related to the
structure of the $\Z_6$ equations in section~4. There was a non--zero
trace for the $\gamma$ matrices, corresponding to some number of
branes which could not be grouped in such a way as to define an object
uncharged under the twisted R-R six--forms. Notice also that there are
8 identical eigenvalues of the $\gamma_{1,5}$ matrix.}.  The first
case gives a factor $U(4)$, while the two ways of doing the latter
give $U(2){\times}U(2)$ and $U(1)^4$ (The minimum number of D5--branes
at any given point is 2, due to the $\Omega$ projection.)

The enhanced gauge symmetries associated to the isolated or coincident
dynamical fivebranes are precisely the structures observed in
ref.\small.  The heterotic fivebrane/small instantons in the dual
heterotic model (expected to exist for each of the models presented
here) is realised in the $\Z_N^A$ orientifold as a family of $2N$
D5--branes forced to move together as one when away from fixed points.

There are disconnected families of models formed by distributing the
D5--branes among different fixed points in a $\ZN$ invariant way.
Once again (if possible) complete fivebranes can move off the fixed
points, giving $SU(2)$ and $USp(m)$ factors from the bulk as usual,
while leaving behind any group of D5--branes which is less than $2N$.
This remainder will produce unitary factors to the gauge group in the
obvious way, with $U(1)$ arising for each isolated pair on a fixed
point, etc.

Let us consider the dynamics of fivebranes a little more, in the light
of the fixed point structure of the orbifolds.  D5--branes may move off
fixed points in groups of $2N$. This is accomplished by recalling the
transformation properties of the fixed points under the $\ZN$
generator, discussed in section~2. Taking the most complicated
example, $K3({\bf Z}_6)$, we see that twelve D5--branes may move off
the ${\bf Z}_6$ fixed point together. Also allowed is for six
D5--branes to move off a ${\bf Z}_3$ fixed point at the same time as
another six move off its doublet partner.  Similarly, a triplet of
${\bf Z}_2$ fixed points can each eject four D5--branes
simultaneously. These are all of the ways in which a dynamical
fivebrane may venture into open space, away from the ALE singularities
of the $K3({\bf Z}_6)$ orbifold.

The $K3({\bf Z}_4)$ example is more straightforward. A ${\bf Z}_4$ ALE
point can yield eight D5--branes, the basic dynamical unit in this
example.  Alternatively, one of the six pairs of ${\bf Z}_2$ fixed
points can produce the same type of object.

The 55 sector of the $K3({\bf Z}_3)$, as discussed before, is trivial:
There are no D5--branes.

D5--branes in the bulk (away from fixed points) are only subject to
the $\Omega$ projection.  The $2N$ components of the dynamical
fivebrane, have split into $N$ pairs, one in each sector of the local
division of ${\bf R}^4$ into slices performed by the rotation
generated by $\ZN$ The orbifold ensures that their movements are
correlated, however, as $\alpha_N$ relates each sector. For example
for $K3({\bf Z}_6)$, we have the following  picture:

\vskip1.0cm
\hskip3.5cm\epsfxsize=1.5in\epsfbox{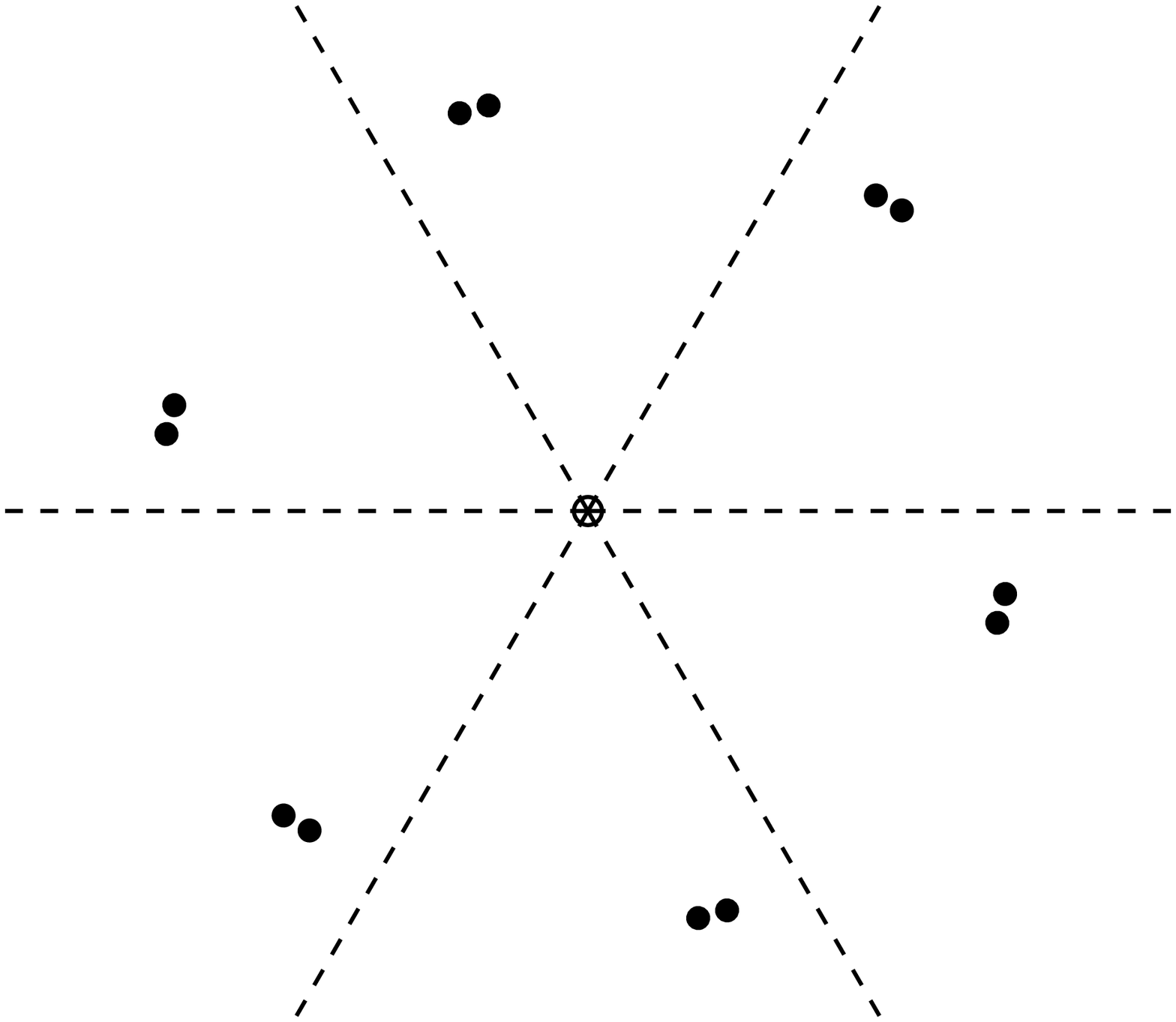}

where the central point is either a ${\bf Z}_6$ ALE singularity, or a
doublet of ${\bf Z}_3$ ALE singularities, or a triplet of ${\bf Z}_2$ ALE
points. In the latter two cases, those points have been identified in
order to form the ${\bf Z}_6$--invariant fundamental domain. The pair
of points\foot{Of course, we have only drawn one of the two complex
planes of the original $T^4$ here. Furthermore, the worldvolume of the
D5--branes under discussion here fill out the non--compact six
dimensional spacetime.}\ is a D5--brane pair, forced to move together
by the orientifold\ericjoe.

For the $\Z_6^B$ model, the only non--trivial $B$--type model with
D5--branes, we have a slightly different situation. The spectrum is
the same (by T--duality) as the model $\Z_3^A$ where the gauge group
is carried by D9--branes. Here, however, the gauge group is carried by
D5--branes on a single fixed point. We can connect to other models by
Higgsing, moving dynamical fivebranes off the fixed point as before.
Also as before, we can make families in disconnected sectors of moduli
space by making different starting configurations of distributions of
D5--branes. 

However, there are some important differences. First, the number of
D5--branes making up one fivebrane is 6. This can be seen easily from
the fact that (as mentioned before) the spacetime orbifold group is
not $\Z_N$, but $\Z_{N/2}$ for the $\Z_N^B$ orientifold group.
Therefore the D5--branes are forced to move as triplets by the $\Z_3$ and
these triplets are forced to move in correlated pairs by the orientifold.

Another difference between $\Z_6^B$ and the $A$--type models is that
the gauge group of a single fivebrane (6 D5--branes) is U(1) and not
SU(2). The reason can be seen as follows: In the $A$--type models the
orientifold group element $\Omega$ correlated the movements of a pair
of D5--branes, and in addition constrained them to be at the same
position on the torus\ericjoe. Here however, the orientifold group
element $\Omega R$ appears instead of $\Omega$ and it correlates the
dynamics of a pair of branes, as before, but they are not coincident,
being mirror images of each other under of the action of $R$. A quick
check of the Higgs mechanism using the spectrum given in the table
shows that indeed $U(1)$ is the correct gauge group for an isolated
dynamical fivebrane in this model, and not $SU(2)$. This is also
consistent with the fact that completely isolated D--branes should
carry $U(1)$ gauge group, and no more.

There are 4 possible dynamical fivebranes in this model.  Starting
with all of the D5--branes on one point, with gauge group
$U(8){\times} SO(16)$ we can move them off in $m$ groups of 6 to give
a breaking to $U(8-2m){\times}SO(16-2m)$, for $m<5$, with factor
$U(1)^m$ if the fivebranes are all isolated, and $U(m)$ if al
coincident. (The intermediate cases are obvious.) Once
$SO(8){\times}U(1)^4$ has been reached corresponding to 8 D5--branes
remaining on the fixed point, and 24 in bulk (forming 4 fivebranes) no
more groups of 6 can be moved off\foot{This can be seen in many
different ways. One way involves noting that (as happened in the
$\Z_6^A$ case, there are eight identical eigenvalues of the
$\gamp_{2,5}$ matrix, corresponding to the number of branes which
cannot be moved off.}.

Let us briefly consider the blow--up limits of the $\Z_N^A$ models
which we have been considering. As we reminded ourselves in section~2,
the smooth $K3$ limit is obtained by performing surgery on the
orbifold limits we have discussed here, where the singular ALE points
are excised and replaced by small ALE gravitational instantons\foot{We
have not considered in any detail the whereabouts of the moduli for
this process. They should exist, and we assume that they do for this
brief discussion.}. It is interesting to consider how some of the
results obtained in the rest of the paper fit in with this process.

Considering first the ${\bf Z}_2^A$ of ref.\ericjoe, we have 32
D5--branes, which are grouped into units of 4 to make 8 dynamical
fivebranes, corresponding to one small instanton on the heterotic
side\small.  The total instanton number (contribution to $\int
F{\wedge} F$) of $K3$ is 24, and so the 16 small ${\cal E}_2$
gravitational instantons from the blow--up must each supply one unit.
Meanwhile, the Euler number (contribution to $\int R{\wedge}R$) of
$K3$ is 24. As the ALE spaces supply all of the curvature for the
smooth manifold, they must each contribute 3/2 to the total.  The
(untwisted) R-R field strength $H_7$ of the R-R six--form $A_6$ (which
naturally couples to the world volume of a fivebrane) must obey total
cancelation of its charge $Q_5\propto (\int F{\wedge}F-\int
R{\wedge}R)$ in the compact internal $K3$ space, which it does with
the given assignments above.  This gives the $H_7$ charge of a single
${\cal E}_2$ space as ${\tilde\mu}^{(2)}_5=1-3/2=-1/2$, measured in
units where we set the dynamical fivebrane's $H_7$--charge to
one\foot{This is not just a natural or convenient choice, but the
physical one: We must not forget Dirac--Teitelboim--Nepomechie
quantisation of the six--form charge. In type IIB, the basic units of
quantisation were carried by D5--branes. Now it is carried by $2N$ of
them. This is possible because the $\Omega$ projection renormalises
the basic unit of charge by 1/2, and the $\Z_N$ projection by another
$1/N$. This generalises the comments made in ref.\ericjoe.}. The
solution of the tadpole equation \tadstwo\ by choosing 32 D5--branes
is simply reinterpreted here as an equation for conservation of
$H_7$--charge: $16{\tilde\mu}^{(2)}_5+8\mu_5=0$.

Let us see how this works in the other models. In the ${\bf Z}_3^A$
case, we learn something simple. Recall that there are no D5--branes
in this case. This means that $H_7$ charge must be zero due to exact
cancelation of the ALE points' instanton number against Euler
number. As there were 9 identical fixed points, we see that the
contribution to Euler number of an ${\cal E}_3$ space must be 8/3, and
their instanton number in these units must be equal and opposite to
this. The $H_7$ charge ${\tilde\mu}^{(3)}_5$ of a ${\cal E}_3$ space is
the difference between these two --- zero --- which is of course why there
are no D5--branes; another interpretation of the absence of D5--branes, as
taught to us by  equation \tadstwo.

The ${\bf Z}_4$ case is the first of the two examples with mixed
species of ALE instanton in the blow--up.  It has 16
singularities to be blown up.  The blow--up replaces 4 ${\bf Z}_4$
points with ${\cal E}_4$ spaces. The other 12 singularities are
paired into 6 doublets and blown up with 6 ${\cal E}_2$'s. We
learned previously that the contribution to Euler number of the ${\cal
E}_2$'s was 3/2 and so we deduce the contribution of an ${\cal E}_4$
to be $15/4$ after simple arithmetic. Consider now instanton
number. We have 32 D5--branes, which clump into groups of 8 to
make 4 dynamical fivebranes, each of which corresponds to one
small heterotic instanton. We also have that one ${\cal E}_2$ space
has instanton number 1. Putting this all together, and doing some more
simple arithmetic, we find that $Q_5=0$ is satisfied when the
instanton number of an ${\cal E}_4$ is $14/4$.  This gives us an $H_7$
charge ${\tilde\mu}^{(4)}_5$ of an ${\cal E}_4$ space equal to $-1/4$.

We turn finally to the ${\bf Z}_6$ case, with its 24 fixed points. We
still have 32 D5--branes, from \tadstwo, and these should correspond
to $32/12$= 2+2/3 dynamical fivebranes/small instantons. So only two
fivebranes can be untethered to fixed points in this model, as we saw
earlier. Moving on, we can put together all that we have learned to
deduce that the combination of one ${\cal E}_6$ space (to resolve the
${\bf Z}_6$ singularity), 4 ${\cal E}_3$ spaces (to resolve the
eight ${\bf Z}_3$ singularities), 5 ${\cal E}_2$ spaces (to resolve
the fifteen ${\bf Z}_3$ singularities) and the 2+2/3 instanton yields
for ${\cal E}_6$ an Euler number contribution of 35/6, and an
instanton number contribution of 34/6.  The $H_7$ charge
${\tilde\mu}^{(6)}_5$ of an ${\cal E}_6$ space is therefore equal to
$-1/6$.

It is quite pleasing to observe the results of the above numerical
exercise. The fact that for an ${\cal E}_m$ ALE space, the $H_7$
charge is given by ${\tilde\mu}^{(m)}_5=-1/m$ for $m$ even and is zero
for $m$ odd is indeed what we deduced from studying and solving the
tadpole equations. This also means that we can place $1/m$ of a
fivebrane on each ${\cal E}_m$ space to cancel all $H_7$--charge {\sl
locally}, in any of the models\foot{This is a configuration which will
be of some interest to us later.}. As we saw above, we learned that
this was possible from direct computation of the spectrum.

What we have done here is simply to rephrase the results we obtained
in a language quite well--suited to the discussion of dual six
dimensional heterotic string theory compactifications.  It will be
exciting to return to more of these issues in some detail.

\newsec{Anomalies}
We have verified that the irreducible $\tr R^4$ terms in the
gravitational anomaly polynomials vanish for these spectra, by
checking that the equation
\eqn\anoms{n_H-n_V=244-29n_T} is satisfied. (Here, $n_H,n_V$ and
$n_T+1$ are respectively 
the numbers of hypermultiplets, vector multiplets and
tensor multiplets in the six dimensional supergravity model.)
We have also checked that the irreducible $\tr F^4$ terms in the gauge anomaly
polynomials vanish.

That these anomalies cancel is a fine example of the interplay between
the $K3$ geometry, the open string sectors, and closed string sectors.
In the $\Z_2^A$ example of ref.\ericjoe, the number $n_T$
is zero, and the 20 closed string hypermultiplets contribute to the
cancelation of the anomaly in the usual way. One might have expected
that this procedure would have persisted, the closed string sector
providing the 20 hypermultiplets as usual, and the open string
contribution to the vector multiplets and hypermultiplets changing in
such a way as to make sure that the anomaly is canceled.

A first realisation that this expectation will go spectacularly wrong
is the observation that the $Z_4^B$ model has no open string sectors:
it is a purely closed string theory. So the contribution of the open
string vectors and hypermultiplets is absent, and the resulting model
would be terribly sick due to gravitational anomalies. However, it is
seen that the number of closed string hypermultiplets reduces to 12, in
exchange for 8 extra tensor multiplets, rendering the theory anomaly
free, and producing a consistent closed unoriented string
theory\foot{This model is the same as the closed string sector of
ref.\atish, obtained while studying a different orientifold group.}.

The same sort of thing happens for the other models, with a varying
split of the number 20 between the number of hypermultiplets and extra
tensor multiplets, in just such a way as to combine with the open
string spectrum to give an anomaly free theory.

For the mixed anomalies, we expect that the anomalies all factorise in
a way consistent with cancelation by a generalisation\sagnotti\ of the
Green--Schwarz mechanism. This was studied for the
$\Z_2^A$ model in ref.\joetc, and for a related model (similar to
$B$--type models) in ref.\atish.  Similar structures will arise here,
and we shall present a discussion of these and related issues in a
separate publication\moretocome.

\newsec{Conclusions}

We have investigated a large family of orientifold models,
corresponding to consistent open and closed string unoriented string
theories propagating on the $K3$ surface, yielding six dimensional string
theories with ${\cal N}{=}1$ spacetime supersymmetry.

Along the way we have observed a truly remarkable interplay between
three separate (but related) elements: The local behaviour of open
string sectors (D--branes) in the neighbourhood of ALE points\foot{See
ref.\mikegreg\ for a complementary study of D--branes and ALE
singularities.}, as captured by the tadpole equations; the behaviour
of the unoriented closed string sectors; and the geometry of the $K3$
manifold.  In every case we studied, the geometry of $K3$ seemed to
know exactly  how to combine the 32 D9-- and D5--branes with
 the closed string twisted
sectors in such a way as to give a consistent model.

This is further confirmation that the details of orientifold
construction set out in ref.\ericjoe\ are a powerful addition to (and
refinement of) the tools available to study ${\cal N}{=}1$ string
models, reuniting open and unoriented strings with heterotic strings
in the rich family of models which should be studied in this context.

The observation of refs.\refs{\small,\ericjoe}\ that the type~I string
theory's basic dynamical fivebranes have a symplectic projection on
their Chan--Paton factors was seen to persist in these models, and is
expected to do so as long as the D5--branes are away from special
points. The mechanism by which such a dynamical fivebrane arose in
each model is easily traced to the spacetime and orientifold
symmetries.  Further to this, in each particular model, we saw that
the individual ALE instanton/fixed point structures give rise to new
families of enhanced gauge symmetry groups, with an associated
spectrum of charged matter.

We expect that in the case of the even $N$ $\ZN^A$ models, there will
be an interpretation in terms of some dual families of $SO(32)$
heterotic string models compactified on $K3$, where the D9--brane
contributions will be perturbatively realised, while the D5--brane
contributions will appear as non--perturbative symmetries. This will
generalise the ideas in refs.\refs{\small,\ericjoe,\joetc,\hethet}. It
will be certainly interesting to investigate these models further\moretocome.

\bigskip
\medskip

\noindent
{\bf Acknowledgments:}

\noindent
We would like to thank Shyamoli Chaudhuri for useful conversations and
Joe Polchinski for encouragement and many illuminating discussions.
CVJ would also like to thank Beth Ellen Rosenbaum and Al Shapere for
their kind hospitality while this paper was being completed.  This
work was supported in part by the National Science Foundation under
Grants PHY91--16964 and PHY94--07194.
\listrefs
\bye